\documentclass[acmsmall]{acmart}

\usepackage{acronym} 
\usepackage{algorithm} 
\usepackage{algorithmic}
\usepackage{booktabs}
\usepackage{clrscode3e}

\usepackage{multirow}
\usepackage{graphicx}
\usepackage{diagbox}
\usepackage{textcomp}
\usepackage{xcolor}
\usepackage{tabularx}
\usepackage{threeparttable}
\usepackage{adjustbox}
\usepackage{tabularx}
\usepackage{tcolorbox}  
\usepackage{subcaption}  
\usepackage{caption}   
\usepackage{wrapfig}
\usepackage{enumitem}

\AtBeginDocument{%
  \providecommand\BibTeX{{%
    \normalfont B\kern-0.5em{\scshape i\kern-0.25em b}\kern-0.8em\TeX}}}

\setcopyright{acmcopyright}
\copyrightyear{2025}
\acmYear{2025}
\acmDOI{XXXXXXX.XXXXXXX}

\acmJournal{TOIS}
\acmVolume{00}
\acmNumber{0}
\acmArticle{000}
\acmMonth{0}

\begin{document}

\title{An Efficient  LLM-based  Evolutional Recommendation with Locate-Forget-Update Paradigm}



\author{Hao Liu}
\email{haoliu0723@gmail.com}
\affiliation{
  \institution{Hefei University of Technology}
  \city{Hefei}
  \state{Anhui}
  \country{China}
  \postcode{230009}
}
\author{Le Wu}
\authornote{Corresponding Author.}
\email{lewu.ustc@gmail.com}
\affiliation{
  \institution{Hefei University of Technology}
  \city{Hefei}
  \state{Anhui}
  \country{China}
  \postcode{230009}
}
\author{Min Hou}
\authornotemark[1]
\email{hmhoumin@gmail.com}
\affiliation{
  \institution{Hefei University of Technology}
  \city{Hefei}
  \state{Anhui}
  \country{China}
  \postcode{230009}
}
\author{Han Wu}
\email{ustcwuhan@gmail.com}
\affiliation{
  \institution{Hefei University of Technology}
  \city{Hefei}
  \state{Anhui}
  \country{China}
  \postcode{230009}
}
\author{Kun Zhang}
\email{zhang1028kun@gmail.com}
\affiliation{
  \institution{Hefei University of Technology}
  \city{Hefei}
  \state{Anhui}
  \country{China}
  \postcode{230009}
}
\author{Xin Li}
\email{leexin@ustc.edu.cn}
\affiliation{
  \institution{IFLYTEK Research}
  \city{Hefei}
  \state{Anhui}
  \country{China}
  \postcode{230088}
}
\author{Si Wei}
\email{siwei@iflytek.com}
\affiliation{
  \institution{IFLYTEK Research}
  \city{Hefei}
  \state{Anhui}
  \country{China}
  \postcode{230088}
}

\renewcommand{\shortauthors}{Liu et al.}
\begin{abstract}
Nowadays, Large Language Models (LLMs) have shown exceptional performance in sequential recommendations, and the adoption of LLM-based recommender systems (LLMRec) is becoming increasingly widespread in existing e-commerce platforms. 
Despite the impressive performance, the constant high volume of new user-item interactions makes it difficult to adapt to the evolution of user preference over time, especially for LLM-based recommender systems. 
The challenge arises from the large number of parameters in LLMs, which makes traditional evolution methods (i.e., Re-training or Fine-tuning) impractical. Specifically, Re-training with all interactions results in prohibitively high computational costs. On the other hand, fine-tuning with only new interactions leads to preference forgetting among inactive users, ultimately compromising overall performance. 

To tackle this problem, we propose EvoRec, an efficient Locate-Forget-Update framework designed for LLM-based recommender systems to model the evolution of user preferences. EvoRec identifies a small set of parameters associated with preference changes and updates them precisely, thereby saving computational resources while maintaining strong recommendation performance.
Specifically, we first \textit{Locate} the sensitive layers by monitoring hidden states when the model processes paired inputs of original and updated interaction sequences.
After identifying these layers, we \textit{Forget} ineffective outdated user interactions, which enhances the effectiveness of the sequences.
Finally, we \textit{Update} the parameters of the located sensitive layers using the filtered user interactions to evolve user preferences.
Notably, the modified parameters account for only $30\%$ of LoRA adapter parameters, with no additional parameters introduced.
Extensive experiments on two real-world datasets demonstrate that, compared to existing methods, EvoRec not only efficiently evolves LLMRec to adapt to the preferences of active users, but also preserves the interests of inactive users from being disturbed during evolution. Codes are available at \url{https://github.com/Tam-JQK/EvoRec}.
\end{abstract}

\maketitle


\section{Introduction}
\begin{figure}[ht]
    \centering
    \includegraphics[width=\textwidth]{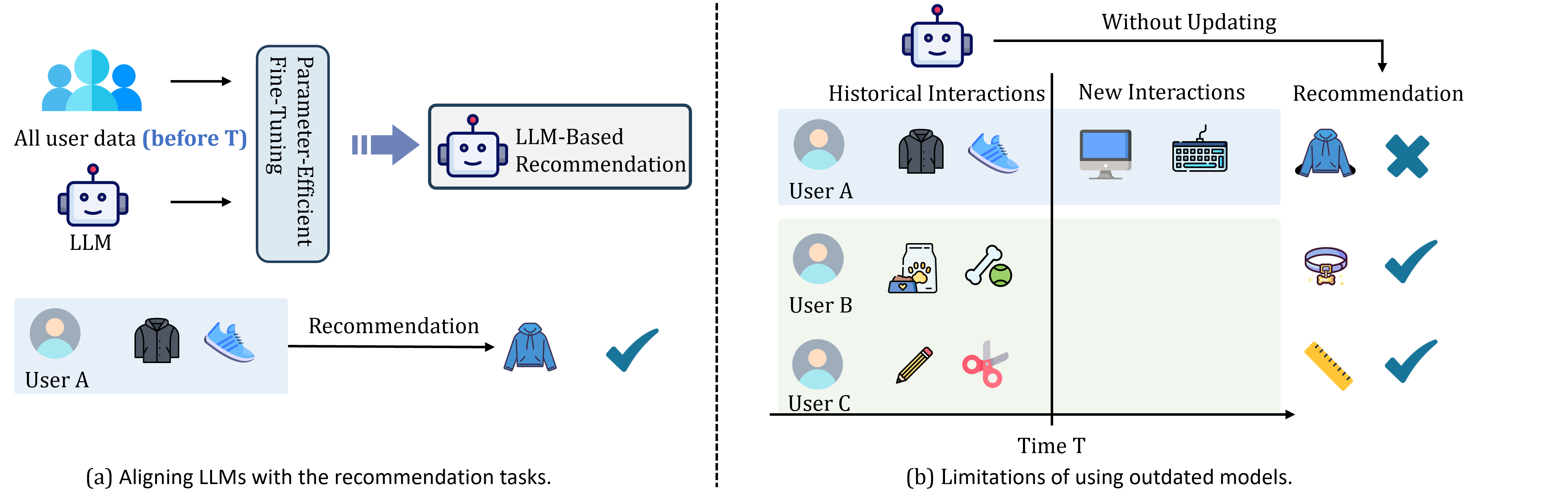}
\caption{Illustration of preference drift challenge in personalized recommendation. (a) By applying efficient fine-tuning techniques to all user data before time T, LLMs achieve high recommendation performance at the current stage. (b) personalized recommendation systems face a fundamental challenge: user preferences are not static but evolve over time. As user preference drift (e.g., User A's preference changes from clothing to electronics), existing LLM-based recommenders struggle to adapt and provide accurate personalization. In contrast, for users whose preferences remain stable (e.g., Users B and C), the models can still deliver reliable recommendations.
}
    \label{fig:introduce_1}
\end{figure}
\noindent Based on historical interactions, sequential recommendation~\cite{seq_1,seq_2,seq_3,seq_4,liao_1,hu_1,dang_1} aims to capture users' interests and predict the next item they are most likely to engage with.
Over time, various methods have been proposed to improve the performance of sequential recommendations. Early studies utilize matrix factorization (e.g., FPMC~\cite{FPMC}) and sequence neural networks (e.g., GRU4Rec~\cite{GRU4Rec}, Bert4Rec~\cite{bert4Rec}), etc. More recently, Large Language Models (LLMs)\cite{qwen3,deepseek,kimi,longlife_1,longlife_2,longlife_3} have demonstrated strong capabilities in world knowledge comprehension and complex reasoning, which has spurred increasing interest in applying them to recommendation tasks\cite{ilora,bao2023tallrec,zheng2024harnessing,liao2024llara}. These methods take users' sequential behavior as a list of item descriptions, and use
parameter efficient fine-tuning techniques for generative recommendation. For example, TALLRec~\cite{bao2023tallrec} leverages LoRA fine-tuning to efficiently align large language models with recommendation tasks. Due to the superior performance, LLM-based recommendation is a compelling direction for future exploration.

\begin{figure}[t]
    \centering
    \includegraphics[width=0.95\textwidth]{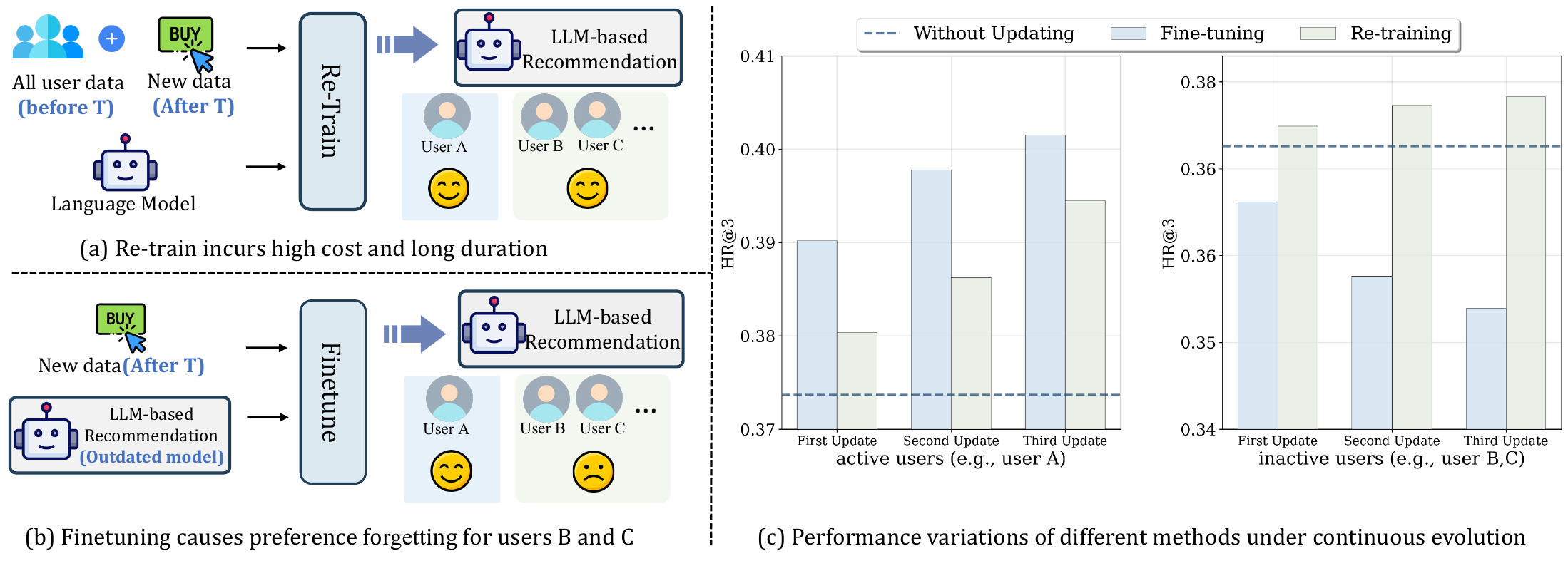}
    \caption{Limitations of two common evolution paradigms: (a) Re-training with new data preserves performance for both evolving (user A) and stable (users B, C) preferences but incurs high cost; (b) Finetuning adapts to user A but causes preference forgetting for users B, C; (c) Results of three consecutive finetuning rounds further confirm this preference forgetting phenomenon.}
    \label{fig:introduce_2}
\end{figure}

Despite the impressive performance of LLMs in recommendation tasks, they still face limitations when handling newly generated user interactions. As illustrated in Figure~\ref{fig:introduce_1}, active users’ preferences continuously evolve over time (e.g.,user A experiences a preference drift from clothing to electronics), while inactive users (e.g., users B and C) maintain stable and unchanged preferences. Such partial preference drift pose significant challenges for LLMRec. Using outdated LLMRec inevitably leads to inaccurate recommendations for active users with evolving interests (e.g., user A). Another critical issue is that, unlike traditional lightweight recommendation models, re-training LLMs is prohibitively expensive in terms of time and computational resources due to their massive parameter size~\cite{zhang2020retrain}. Fine-Tuning strategies, such as LoRA~\cite{hu2021lora}, provide a practical solution. By tuning only a small portion of parameters with new data from active users, LLMRec can quickly adapt to their latest preferences without the prohibitive cost of full re-training. Although this can improve the recommendation performance for active users, it largely leads to preference forgetting for inactive users (e.g., users B and C). As shown in Figure~\ref{fig:introduce_2}, fine-tuning on post-$T$ interaction data can improve recommendation performance for active users but often causes severe performance degradation for inactive users.
This indicates that existing evolution-oriented methods~\cite{sml,lsta} fail to selectively update the key parameters associated with preference drift, making it difficult to quickly adapt to the evolving interests of active users while preventing preference forgetting for inactive users without recent interactions. For example, LSAT~\cite{lsta} updates user preferences by merging LoRA adapters trained on short-term data, which can easily lead to preference forgetting in scenarios with frequent updates. Therefore, a natural question arises: “In LLMRec, how can we efficiently evolve user preference modeling while ensuring performance for both active and inactive users?”

We argue that the evolution of LLMRec faces two major challenges: the “model level” and the “data level.” At the model level, extensively updating the parameters of the original LLMRec can indeed facilitate the preference evolution of active users, but it also leads to preference forgetting among inactive users, resulting in performance degradation and inefficiency. Thus, how to perform selective updates remains a critical issue. At the data level, not all interactions of active users contribute to preference evolution. Some of them are outdated with respect to current preferences, and using them indiscriminately not only introduces noise but also hinders LLMRec from adapting to users’ latest preferences. Therefore, removing outdated interactions is essential for effective evolution.

To tackle the above problem, in this paper, we develop a novel Locate-Forget-Update framework: EvoRec, to facilitate efficient and effective LLM-based sequential recommendations. 
Specifically, we first develop a novel preference localization module to locate the sensitive layers of LLM-based recommender associated with user preference change. 
Note that this step only involves a small fraction of parameter updating, thereby providing computational efficiency compared to global optimization. 
Then, we employ a lightweight sequential recommendation model to identify outdated interactions from historical interaction sequences and filter out (i.e., forget) these outdated interactions, which can effectively reduce noise for the parameter updating of LLM-based recommender. 
Finally, we use both filtered old interactions and new interactions to precisely update the selected sensitive layers, realizing accurate user preference evolution modeling. 
Moreover, two optimization targets are designed to ensure the learning process. 
Since we only update the selected sensitive layers, not only do the modified parameters account for only $30\%$ of traditional LoRA adapter parameters, but also the performance on inactive users is maintained. 
Extensive experiments over two real-world datasets demonstrate the effectiveness and efficiency of our proposed EvoRec. 

The main contributions of this paper are summarized as follows:
\begin{itemize}
    \item We investigate a novel yet practical research problem: adapting to rapidly evolving preference shifts for active users while ensuring that the preferences of inactive users are not forgotten during the evolution process.
    \item  We propose EvoRec, an evolutionary framework that uses sensitive-parameter localization to quickly identify parameters related to preference shifts, avoiding preference forgetting from large-scale updates. A filtering model is also employed to remove outdated preferences from user interactions, ensuring accurate evolution.
    \item  We conduct extensive experiments on two real-world datasets to demonstrate the effectiveness of our approach.
\end{itemize}

\section{RELATED WORKS}
\subsection{Sequential Recommendation}
The objective of sequential recommendation is to leverage users’ continuous behaviors to predict the next most likely item for interaction. Current methods can be broadly categorized by their network structures, including recurrent neural networks (RNNs)\cite{GRU4Rec,RNN2}, graph neural networks (GNNs)\cite{graph_1,graph_2,graph_3,graph_4}, attention-based networks~\cite{S3-Rec,SASRec,CL4SRec}, and multi-layer perceptrons (MLPs)~\cite{fmlp}. 
Early studies on sequential recommendation primarily rely on recurrent neural networks (RNNs). For example, GRU4Rec~\cite{GRU4Rec} introduces the GRU units to model user sequences, achieving superior performance and practicality compared to Markov chain–based methods~\cite{markov}.
Subsequently, graph neural networks (GNNs) demonstrate stronger capability than RNNs in capturing user interests and modeling item correlations. FGNN~\cite{graph_1} proposed a session-based recommendation model that represents sessions as graphs, formulating next-item prediction as a graph classification task and incorporating a weighted attention layer with a Readout function to capture both sequential and latent item transition patterns. Similarly, SURGE~\cite{graph_2} is a GNN-based model that transforms user behavior sequences into item–item interest graphs, leveraging cluster-aware and query-aware graph convolution with pooling to capture dynamic preferences from noisy data, thereby enhancing performance. Even though GNN-based methods have achieved strong performance, they still face inherent limitations: they rely on multi-hop aggregation to capture user preferences, which makes it difficult to balance accuracy and efficiency. As a result, recent research has largely shifted toward attention-based networks. SASRec~\cite{SASRec} is the first model to introduce self-attention into sequential recommendation, which captures long-term user behavior patterns by focusing on the most relevant past items and achieves significantly better performance. CL4SRec~\cite{CL4SRec}, built on SASRec, incorporates contrastive learning into attention-based sequential recommendation and leverages self-supervised signals derived from data augmentation to learn more robust user representations, thereby alleviating data sparsity. Apart from RNNs, GNNs, and attention-based networks, FMLP~\cite{fmlp}, the first purely MLP-based model for sequential recommendation, is enhanced with learnable frequency-domain filters to mitigate noise in user behavior sequences while leveraging an all-MLP architecture with adaptive filtering. Although the above methods have achieved promising performance, the ID-embedding paradigm faces limitations in adapting to cold-start scenarios and lacks the ability to leverage rich semantic information for enhanced recommendation. The extensive knowledge reserves and strong contextual reasoning capabilities of LLMs provide a novel solution to address this challenge.

\subsection{LLMs for Recommendation}
Due to the strong text comprehension and reasoning capabilities, large language models (LLMs)~\cite{o-lora,kim2024,cmt,tois_1,tois_2} recently gain significant attention in the field of recommendation systems~\cite{llm_survey1,llm_survey2,llm_survey3,guo_3,huang_1,huang_2}. Existing research on LLM-based recommendation tasks can be broadly categorized into two paradigms: LLMs as enhancers~\cite{MoRec,llm_esr,wei2024llmrec,guo_1,guo_2} and LLMs as recommenders~\cite{zheng2024harnessing,liao2024llara,ilora,p5,bao2023tallrec}.
The enhancer paradigm takes advantage of the extensive knowledge of LLMs to extract textual features that enhance the embeddings of traditional models. For example, MORec~\cite{MoRec} preliminarily demonstrates that multimodal information can effectively mitigate cold start and data sparsity issues. Hu et al.\cite{alphafuse} propose AlphaFuse, a language-guided learning strategy that learns ID embeddings in the null space of language embeddings to preserve semantic information. They use singular value decomposition to separate the semantic space and inject collaborative signals into the null space, avoiding semantic degradation and eliminating the need for extra trainable modules. This paradigm typically avoids extensive modification of LLM parameters.
In contrast, the recommender paradigm primarily employs efficient prompt-tuning techniques to align LLMs with recommendation tasks. For example, Bao et al.\cite{bao2023tallrec} utilize instruction tuning to adapt LLMs' extensive knowledge for recommendation while demonstrating strong cross-domain generalization. Liao et al.\cite{liao2024llara} combine traditional recommendation models with Large Language Models to enhance sequential recommendation. It integrates ID-based item embeddings and textual features using hybrid prompting, so to improve the behavioral understanding and world knowledge encoding. Kong et al.\cite{ilora} combine LoRA with a mixture-of-experts (MoE) framework, dynamically adjusting expert parameters through multitask learning, significantly improving recommendation accuracy. Among them, the input typically consists of user-interaction sequences formatted as instructions, while the output is the textual representation or title of the recommended item.
In summary, current LLMRec achieves promising performance in static settings. However, since user preferences may shift over time, relying on outdated LLMRec for recommendations can lead to suboptimal results. Therefore, adapting LLMRec to dynamic user preferences remains a key challenge.

\subsection{Incremental Recommender System}
User preferences naturally evolve over time, and recommendation models that fail to adapt to these changes can become outdated, resulting in suboptimal recommendations. Incremental recommendation~\cite{start1,fine-tuneing1,fine-tuneing2} is designed to continuously evolve recommendation models to capture users’ latest interests, yet most existing studies still focus on traditional recommendation approaches. Existing incremental recommendation methods can broadly be grouped into three categories: Re-training, Fine-tuning, and long–short term preference fusion. (1) Re-training strategies combine newly collected data with historical interactions and re-train the recommendation model from scratch. Beyond full retraining, some studies introduce sampling-based retraining~\cite{sample1,sample2}, which aims to extract the most valuable data from both the original and newly added interactions to update the model more efficiently. For traditional recommendation models with relatively few parameters, such approaches remain feasible; however, for LLM-based recommenders with massive parameter scales, Re-training is computationally prohibitive. (2) Fine-tuning approaches~\cite{fine-tuneing1,fine-tuneing2} incrementally update the model with new data only. Compared with Re-training, this paradigm significantly reduces computational cost and update latency, making it suitable for real-time adaptation. Nevertheless, repeated fine-tuning under continuous updates may lead to preference forgetting, thereby degrading recommendation performance. To alleviate this issue, FT-KL~\cite{FT-KL} introduces a KL-divergence constraint on the model’s logits, preventing the evolved model from deviating excessively from the original one. Similarly, the EWC~\cite{ewc} method measures parameter importance through the Fisher information matrix and protects parameters that store other users’ preferences while adapting to active users, thereby reducing preference forgetting for inactive users. (3) Long–short term preference fusion emphasizes balancing long-term interests with dynamic short-term preferences. For example, LSTTM~\cite{meta1} constructs a global long-term graph and an internal short-term graph, and further applies meta-learning to achieve fast adaptation across different time periods. SML~\cite{sml} introduces a transfer component that transforms the old model into a new one tailored for future recommendations, avoiding the need to reuse historical data while improving both efficiency and accuracy. In LLMRec, LSAT~\cite{lsta} trains low-rank LoRA adapters on recent interactions and searches for an optimal fusion coefficient to combine long-term and short-term modules. Although such strategies reduce the need for large-scale parameter updates, they still face challenges in preventing preference forgetting for inactive users. Different from previous methods, our proposed EvoRec employs a lightweight sequential recommendation model to filter out outdated user interactions. It then precisely locates the key parameters associated with users whose preferences have shifted. Finally, the filtered high-quality sequences are used to update only the identified sensitive parameters, avoiding large-scale parameter modifications that could cause preference forgetting for inactive users, while maintaining strong performance even under continuous updates.

\section{Preliminary}

\subsection{Task Formulation}

We focus on the sequential recommendation task, which aims to predict the future preferences of users based on their historical interaction sequences. Let \(\mathcal{U}\) and \(\mathcal{I}\) denote the sets of users and items, respectively. For a user \(u \in \mathcal{U}\), their interaction history is represented as a chronologically ordered sequence, \(S_u^{\leq T}\), which includes all interactions up to a manually set time \(T\). The sequence is given~by:

\[
S_u^{\leq T} = \left[i_u^1, i_u^2, \ldots, i_u^{T_u}\right],
\]
where \(i_u^k \in \mathcal{I}\) denotes the \(k\)-th interacted item, and \(|L_u^{\leq T}|\) is the sequence length for interactions up to time \(T\).
According to the time \(T\), we divide the user set \(\mathcal{U}\) into two disjoint subsets: \(\mathcal{U}_A\) and \(\mathcal{U}_I\), such that \(\mathcal{U} = \{\mathcal{U}_A \cup \mathcal{U}_I\}\). Here, \(\mathcal{U}_A\) refers to active users whose interactions occur after \(T\), and \(\mathcal{U}_I\) refers to inactive users with no interactions after \(T\).

Then, for a given \( u \in \mathcal{U} \), we represent his/her interaction sequence that occurred before time \( T \) as \( S_u^{\leq T} = \left[i_u^1, \ldots, i_u^{T_u}\right] \). Specifically, for \( u \in \mathcal{U}_A \) with new interactions after \( T \), the complete interaction sequence is given by
\[
S_u^A = \left[i_u^1, i_u^2, \ldots, i_u^{T_u}, i_u^{T_u+1}, \ldots, i_{u}^{T^{'}}\right],
\]
where the total sequence length is \( T^{'} \). 
For better description, we name the set of the first $T_u$ elements as $S_u^{A-}$ and the remaining part as $S_u^{A+}$. Therefore, we have $S_u^A = S_u^{A-} \oplus S_u^{A+}$. Along this line, we acquire two instruction-tuning datasets according to the timestamp \(T\) and the user set:

\[
    \mathcal{D}^{\leq T} = \left\{ (x_u, y_u)^{\leq T} \mid u \in \mathcal{U} \right\}, \quad
    \mathcal{D}_A = \left\{ (x_u, y_u)^{A} \mid u \in \mathcal{U}_A \right\},
\]
where \((x_u, y_u)^{\leq T}\) and \((x_u, y_u)^{A}\) represent the instruction-tuning datasets derived from \(S_u^{\leq T}\) and \(S_u^A\) using the function \(I(\cdot)\), and \(x_u\) and \(y_u\) correspond to textual instructions and their respective responses. The function \(I(\cdot)\) denotes the template for instruction tuning:
\begin{tcolorbox}[title={Instruction-tuning template for LLMs}]
\textbf{Input}: The user’s historical interaction sequence is $[ \text{title}_{24}, \, \text{title}_{15}, \, \text{title}_{28}, \, \dots]$.

\textbf{Output}: The predicted items the user is most likely to interact with in the future are $[\text{1. title}_{9}, \ \text{2. title}_{15}, \  \text{3. title}_{6}].$
\end{tcolorbox}

 The dataset \(\mathcal{D}^{\leq T}\) consisting of all interactions before \(T\), is used to derive an initial LLM-based recommender model $f_{\theta}$. The goal of incremental learning is to efficiently update the model $f_{\theta}$ to $f_{\theta'}$, allowing $f_{\theta'}$ to successfully predict the next interaction item $i_u^{L_k+1}$ given $S_u^A$ for each \(u \in \mathcal{U}_A\), while retaining the ability to predict the next item based on \( S_u^{\leq T} \) for each \( u \in \mathcal{U}_I \).
The process for updating the model is given~below:

\begin{equation}
(\mathcal{D}_A, f_{\theta}) \xrightarrow{update} f_{\theta’}.
\end{equation}

\begin{table}[htbp]
  \centering
  \caption{Summary of key symbols and notations used in this work.}
  \begin{tabular}{ll}
    \toprule
    \textbf{Symbol} & \textbf{Description} \\
    \midrule
    $T$ & The decision time point that separates past and new user interactions. \\
    $\mathcal{U}_A$ & The set of users who have new interactions after time $T$. \\
    $\mathcal{U}_I$ & The set of users whose interactions only occur before time $T$. \\
    $\mathcal{U}$ & The complete set of users, i.e., $\mathcal{U} = \mathcal{U}_A \cup \mathcal{U}_I$. \\
    $\mathcal{I}$ & The set of all items in the system. \\
   $S_u^{\leq T}$ & The interaction sequence of user $u$ before time $T$, where $u \in \mathcal{U}$. \\
    $S_u^{A}$ & The new interactions of user $u$ after time $T$, where $u \in \mathcal{U}_A$. \\
    $S_u^{A-}$ & The filtered sequence after removing outdated interactions from $S_u^{A}$, where $u \in \mathcal{U}_A$. \\
    $S_u^{A+}$ & The interaction sequence of user $u$ after time $T$, where $u \in \mathcal{U}_A$. \\
    $S_u^{'}$ & The final updated interaction sequence of user $u$ for tuning ($S_u^{'} = S_u^{A-} \oplus S_u^{A+}$), where $u \in \mathcal{U}_A$. \\
    $I(\cdot)$ & Function that maps user sequences to instruction fine-tuning templates. \\
    $\mathcal{D}^{\leq T}$ & Instruction-tuning data of users $u \in \mathcal{U}$ before time $T$. \\
    $\mathcal{D}_A$ & Instruction-tuning data of users $u \in \mathcal{U}_A$ after supplementing new interactions. \\
    $H_0^{\leq T}$ & The embedding representation of the pre-$T$ user interaction sequence. \\
    $H_0^{A}$ & The embedding representation of the newly added interactions after $T$. \\
    $\boldsymbol{seq}_u$ & the sequence representation of user $u$ output by filtering model. \\
    $\Phi$ & the set of located sensitive parameters. \\
    $\text{Filter}(\cdot)$ & The filtering function that identifies and removes outdated interactions. \\
    \bottomrule
  \end{tabular}
  \label{tab:symbol}
\end{table}

\subsection{LLM-based Recommender Systems}
The objective of LLMRec is to accurately estimate the relevance score of each candidate item based on the historical interactions of the user, ultimately producing a ranked list of recommended items in descending order of predicted relevance. Given that LLMs typically consist of billions of parameters, fully updating the entire model during adaptation is computationally prohibitive. Inspired by recent advancements in efficient parameter tuning techniques~\cite{AdaLoRA,adapater}, most LLM-based recommendation methods use the LoRA~\cite{hu2021lora} (Low Rank Adaptation) framework to insert adapters into the LLM, aligning the original model with the recommendation task. Let 
$\theta_{\text{pre}}$ represents the frozen parameters of the original LLM, and $\Delta\Theta$ denotes the trainable parameters of the LoRA. During the fine-tuning phase, only $\Delta\theta$ is updated, while the other parameters are frozen. This approach significantly reduces computational costs and training time.
We define an instruction-tuning dataset
$\mathcal{D} = \{(x_i, y_i)\}_{i=1,\ldots,|\mathcal{D}|}$. To better align the LLM with the recommendation task, we minimize the negative log-likelihood of generating the target recommended items based on the historical interactions of the user:
\begin{equation}
    \max_{\Delta\theta}\sum_{\left(x_{i}, y_{i}\right) \in \mathcal{D}} \sum_{t=1}^{|y_i|} \log \left(P_{\theta_{\text{pre}} + \Delta\theta}\left(y_{i, t} \mid x_{i}, y_{i,<t}\right)\right).
    \label{sft}
\end{equation}

The core of LoRA is to model the weight adjustment $\Delta W$ of each target layer as the product of two learnable low-rank matrices:
\begin{equation}
    \Delta W = \mathbf{B}\mathbf{A},
    \label{chazhi}
\end{equation}
where $\mathbf{B} \in \mathbb{R}^{d_{\text{out}} \times r}$ and $\mathbf{A} \in \mathbb{R}^{r \times d_{\text{in}}}$. Here, $r$ is a tunable rank parameter that is significantly smaller than both $d_{\text{in}}$ and $d_{\text{out}}$, enabling efficient adaptation to recommendation tasks. The pre-trained weights $\theta_{\text{pre}}$ remain frozen during this process, ensuring that the computational and storage costs are minimal.

\begin{figure*}[t]
    \centering \includegraphics[width=1\textwidth]{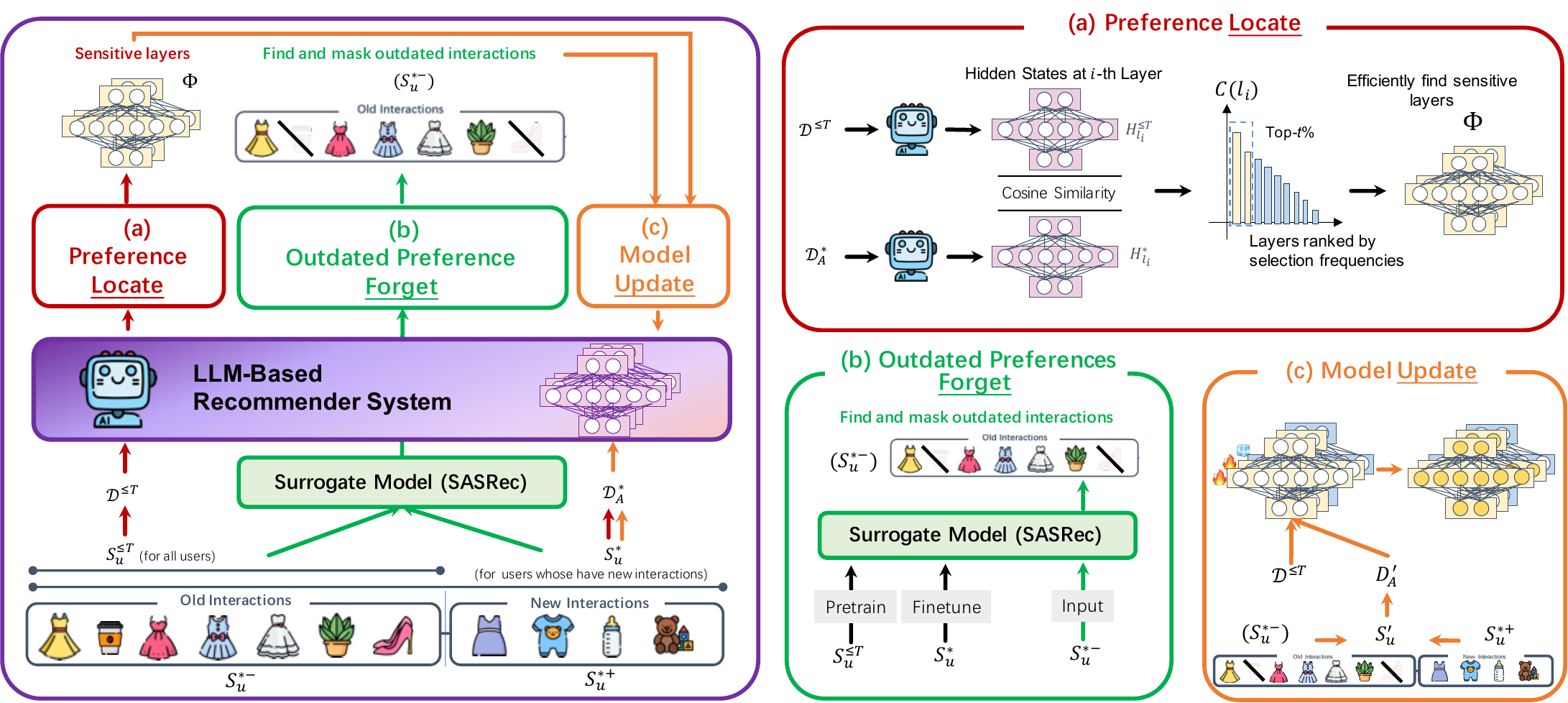}
    \caption{Illustration of our proposed EvoRec framework. First, we input both the historical and recent interaction sequences of users into the outdated LLM-Based Recommender System. By tracking the differences in their hidden states, we identify the sensitive parameters that exhibit the greatest divergence between past and updated preferences. Second, we employ a filtering model to filter out outdated interactions based on relevance scores. Finally, EvoRec updates the identified parameters using the filtered data, ensuring adaptation to users' latest preferences.}
    \label{fig:remove_out}
\end{figure*}
\section{Methodology}

\subsection{Overview}
\begin{figure*}[t]
    \centering \includegraphics[width=1\textwidth]{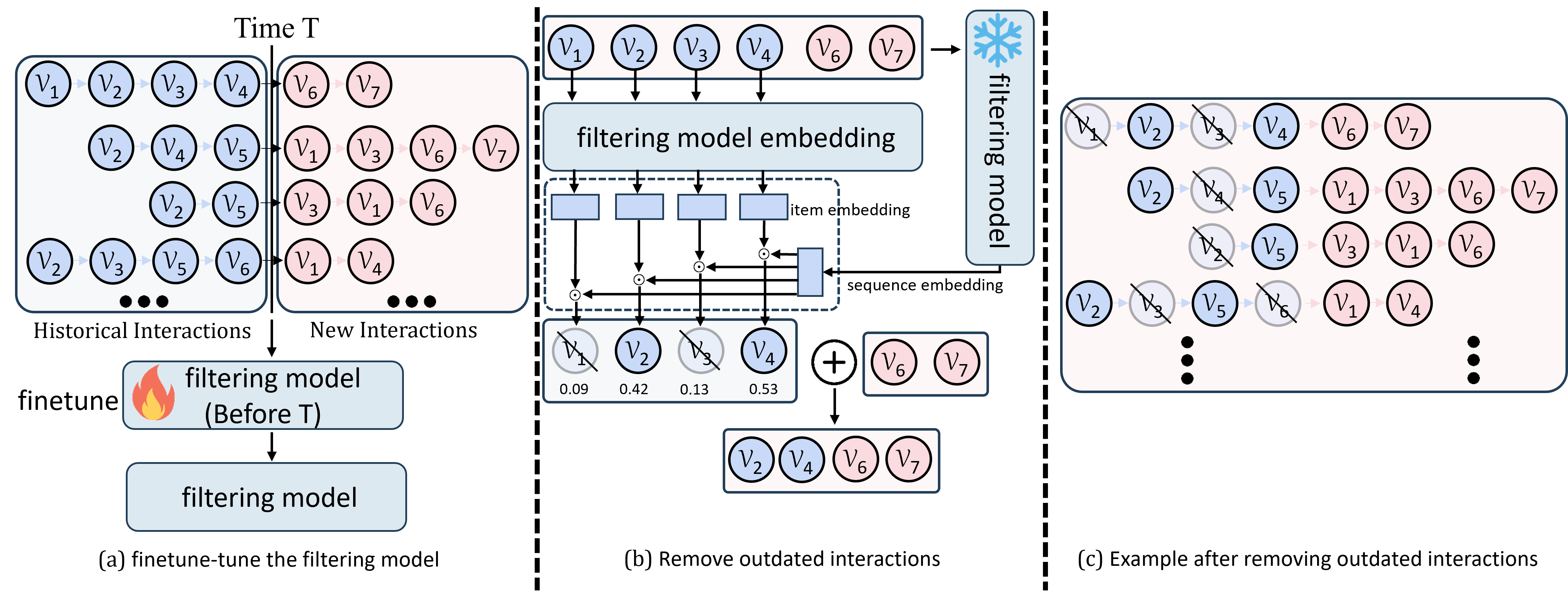}
    \caption{Illustration of the outdated interaction removal process. (a) Fine-tuning the filtering model with recent data to quickly adapt to users’ latest preferences. (b) Removing interactions with low contribution scores based on their impact on user sequence modeling (considered outdated). (c) Example after removing outdated interactions.}
    \label{fig:framework}
\end{figure*}

In this section, we propose EvoRec, an incremental learning framework for LLM-based recommendation, built on the Locate-Forget-Update paradigm. As shown in Figure~\ref{fig:framework}, the EvoRec update process consists of three main steps. First, we input the past and recent interaction sequences in $\mathcal{U}_A$ into an initial LLMRec (LLM-based recommendation methods) trained on $\mathcal{D}^{\leq T}$ before time $T$. We monitor hidden states across different layers to identify the sensitive parameters that contribute to shifts in user preferences (Section 3.2). Second, we employ a filtering model to filter out outdated interactions based on relevance scores, effectively reducing their negative impact on sequential modeling (Section 3.3). Finally, EvoRec updates only the sensitive parameters identified using the filtered interaction data to refine user preferences (Section 3.4). By applying EvoRec, we significantly improve the user experience in $\mathcal{U}_A$ while largely mitigating the forgetting of preferences in $\mathcal{U}$, ensuring that the overall performance of LLMRec remains unaffected.
\subsection{Locate Preference}
Most incremental recommendation methods involve large-scale parameter updates when evolving the model. While this allows LLMRec to quickly adapt to the updated preferences of active users, such extensive updates can also undermine its recommendation performance for inactive users. To adapt the LLMRec in a way that precisely updates the key parameters for the evolving preferences of active users while preserving the parameters encoding the interests of inactive users, we focus on identifying sensitive parameters that capture the differences between user historical and updated interactions. Specifically, given the truncated interaction sequence $S_u^{\leq T}$ for user $u \in \mathcal{U}_A$ and the whole interaction sequence $S_u^A$ for user $u \in \mathcal{U}_A$ with new interactions, the embedding layer $E$ is first applied to obtain their respective token embeddings:

\begin{equation}
H_0^{\leq T} = E(I(S_u^{\leq T})), \quad H_0^{A} = E(I(S_u^{A})).
\end{equation}
Here, $I(\cdot)$ denotes inserting the user sequence into the Instruction-tuning template for LLMs. These embedded representations serve as the initial hidden states and are then propagated through \( L \)-layer transformer blocks, whose layers are denoted as \( \{\ell_1, \ell_2, \dots, \ell_{i}, \dots, \ell_L\} \). 
At the $i$-th layer, the hidden states of interaction sequences are denoted as $H_{\ell_i}$, evolving according to:
\begin{equation}
H_{\ell_i} = H_{\ell_{i-1}} + \text{MLP}_{\ell_{i-1}}(H_{\ell_{i-1}}) + \text{Att}_{\ell_{i-1}}(H_{\ell_{i-1}}),
\end{equation}
where \( \text{MLP}_{\ell_{i-1}}\) and \(\text{Att}_{\ell_{i-1}}\) represent the multi-layer perceptron and attention mechanisms at layer \(\ell_{i-1}\), respectively.

For each pair of sequences \((S_u^{\leq T}, S_u^{A})\), we will acquire \(H_{\ell_i}^{\leq T}\) and \(H_{\ell_i}^{A}\) at the $i$-th layer. Then, we employ cosine similarity to measure the differences between hidden states:
\begin{equation}
s_{\ell_{i}} = \frac{\langle H_{\ell_{i}}^{A}, H_{\ell_{i}}^{\leq T} \rangle}{\|H_{\ell_{i}}^{A}\|_2 \cdot \|H_{\ell_{i}}^{\leq T}\|_2}.
\end{equation}

In each data pair, the most sensitive layer, denoted \(\ell_{\text{sensitive}}\), is determined using the following formula:
\begin{equation}
    \ell_{\text{sensitive}} = \underset{\ell_{i} \in \{\ell_1, \ell_2, \ldots, \ell_L\}}{\arg\min} s_{\ell_{i}}.
\end{equation}

After processing all pairs of historical and new interaction sequences, we track the selection frequency $C(\ell_i)$ for $\ell_i$, which indicates how often a layer is chosen as the most sensitive in all samples. The calculation of $C(\ell_i)$ is as follows:
\begin{equation}
C(\ell_i) = \sum_{j=1}^{|\mathcal{U}_A|} \mathbf{1}(\ell_i \text{ as } \ell_{\text{sensitive}} \text{ by the } j\text{-th pair of instances}),
\end{equation}
where \(|\mathcal{U}_A|\) is the total number of $\mathcal{U}_A$, and \(\mathbf{1}(\cdot)\) is an indicator function that returns 1 if layer \(\ell_i\) is selected at the \(j\)-th pair of instances, and 0 otherwise.

Finally, we define the set of globally sensitive layers \(\Phi\) based on a predefined threshold \(t\), which represents the top \(t\%\) of layers with the highest selection frequencies:
\begin{equation}
\Phi = \text{Top-}t\%(C(\ell_1), C(\ell_2), \dots, C(\ell_L)).
\end{equation}
Here, the layers in $\Phi$ are considered significantly affected by changes in user preferences and are therefore prioritized for updating.




\subsection{Forget Outdated Preference}
For active users, interactions after the time \(T\) reflect short-term interests, while earlier interactions represent their long-term preferences. Although these long-term interactions occurred relatively earlier, they still play an important role in enabling LLMRec to deeply understand user preferences. However, only a subset of long-term interactions contributes positively to evolving user preferences, while the remaining outdated interactions are irrelevant to users’ latest preferences. Moreover, existing research~\cite{zheng2024harnessing} has demonstrated that overly long sequences can negatively affect the ability of LLMRec to model user behaviors. Therefore, EvoRec is designed to retain interactions that are useful for evolving preferences while discarding the outdated ones. To mitigate the negative impact of outdated items in user interactions, we employ a filtering model, e.g., SASRec~\cite{SASRec}, to filter out these items (see Figure~\ref{fig:remove_out} for an example of this process). 
Specifically, the filtering model is first trained on data $S_u^{\leq T}$, i.e., the interactions before time $T$, using the same user interactions as LLMRec but in an independent and parallel manner. Then, to adapt to the new interactions in $\mathcal{U}_A$, we fine-tune the model on the entire interaction sequence $S_u^{A}$, focusing on users whose preferences have shifted so that it quickly adapts to their new preferences.
The detailed process is as follows:
\begin{equation}
\boldsymbol{seq}_u = \text{SASRec}(S_u^{A}),
\end{equation}
where \(\boldsymbol{seq}_u\) represents the sequence representation of user \(u\). We compute the user's preference score for item \(i\) at step \((t+1)\) based on the historical interactions as follows:
\begin{equation}
P(i_{t+1} = i \mid i_{1:t}) = \mathbf{e}_i^\top \boldsymbol{seq}_{u},
\end{equation}
where \(\mathbf{e}_i\) is the embedding representation of item \(i\). We adopt the Bayesian Personalized Ranking (BPR) loss~\cite{BPR_MF} to optimize the model parameters:
\begin{equation}
L = - \sum_{u \in \mathcal{U}_A} \sum_{t=1}^{n} \log \sigma \left( P(i_{t+1} \mid i_{1:t}) - P(\bar{i}_{t+1} \mid i_{1:t}) \right),
\end{equation}
where each ground-truth item \(i_{t+1}\) is paired with a randomly sampled negative item \(\bar{i}_{t+1}\).

Next, to filter out outdated items, we calculate the scores of each item based on its final sequence representation. Let \( r_{u} \) denotes the relevance score vector of all previous interactions for the user \( u \). The score \( r_{u} \) is computed using the following formula:  
\begin{align}
& r_{u} = \text{SASRec}_{\text{emb}}\left(S_u^{A-} \right) \cdot \boldsymbol{seq}_{u},
\end{align}
where \( \text{SASRec}_{\text{emb}} \) represents the embedding layer of the filtering model. The filtering model is a low-dimensional attention network with only one or two layers, and it operates independently of LLMRec with minimal additional overhead.
After obtaining the relevance scores for all past interactions of the user \( u \), we filter out the bottom \( K \) interactions with the lowest scores. Specifically, we apply a filtering function to remove these interactions from \( S_u^{A-} \), denoted as \( (S_u^{A-}) = \text{Filter}(S_u^{A-}) \). The updated user sequence after filtering is then represented as \( S_u^{'} = (S_u^{A-}) \oplus S_u^{A+} \). Modeling with this refined sequence enables a more efficient capture of short-term user preferences while mitigating the influence of obsolete interactions.

 \subsection{Update Model}

 The selection of parameters and data is crucial during the LLMRec update process to adapt to the evolving preferences of active users. Therefore, we employ a sensitive-parameter localization module to identify the key parameters associated with user interest shifts, thereby preventing interference with inactive users. At the same time, a filtering model is used to refine the user interaction sequence. Given the large number of parameters in LLMs, we use the filtered sequence $S_u^{'}$ to update the sensitive parameters that have been identified. Specifically, after time $T$, some users engage in new interactions, suggesting that the preferences previously captured by the model are likely to have shifted. Thus, the goal of our update is to enhance the model's ability to predict next item based on $S_u^{'}$, while mitigating the negative impact of potential conflicts between old and new preferences.
The preference alignment loss is defined as:
\begin{equation}
   \mathcal{L}_e = -\min_{\Phi}\sum_{\left(x_{i}, y_{i}\right) \in \mathcal{D}_A^{'}} \sum_{t=1}^{|y_i|} \log \left(P_{\Theta_{\text{pre}} + \Delta\Theta}\left(y_{i, t} \mid x_{i}, y_{i,<t}\right)\right),
    \label{sft}
\end{equation}
where \( \mathcal{D}_A^{'} \) is the instruction tuning data set obtained by applying the function \( I(\cdot) \) to \( S_u^{'} \). To ensure that the preference update causes minimal preference forgetting for the users in the set \( \mathcal{U}_I \), where \( \mathcal{U}_I = \mathcal{U} \setminus \mathcal{U}_A \), we introduce a consistency loss:
\begin{equation}
    \mathcal{L}_c = \sum_{u \in \mathcal{U}_I} \mathrm{KL} \left( f_{\theta’}(I(S_u^{\leq T})) \, \middle\| \, f_{\theta}(I(S_u^{\leq T})) \right).
\end{equation}
The total loss for optimizing user preference alignment is given by:
\begin{equation}
    \mathcal{L}_{\text{total}} = \mathcal{L}_e + \lambda 
 \mathcal{L}_c,
\end{equation}
where $\lambda$ is a hyperparameter balancing the alignment and consistency losses.

By minimizing $\mathcal{L}_{\text{total}}$, we iteratively update the sensitive parameters as follows:
\begin{equation}
    \begin{array}{l}
\Phi^{t+1}=\left[\Phi_{1}^{t+1}, \cdots, \Phi_{\ell_{\text {sensitive }}}^{t+1}, \cdots, \Phi_{L}^{t+1}\right] \\
=\left[\Phi_{1}^{t}, \cdots, \Phi_{\ell_{\text {sensitive }}}^{t}-\nabla_{\Phi_{\ell_{\text {sensitive }}}^{}} \mathcal{L}_{\text {total }}, \cdots \Phi_{L}^{t}\right],
\end{array}
\end{equation}
where $\ell_{\text{sensitive}}$ is the layer identified as most sensitive to the difference between old and new preferences, $\nabla_{\Phi_{\ell_{\text {sensitive }}}^{}}$ is the gradient for $\Phi_{\ell_{\text {sensitive }}}$. This update ensures that the model effectively captures the evolving preferences of the user while preserving its overall robustness and generalizability.

\begin{algorithm}[t]
\footnotesize
\caption{The EvoRec update process (Locate -- Forget -- Update).}
\label{alg:evorec}
\begin{algorithmic}[1]
\REQUIRE Pretrained LLMRec $f_{\theta}$ (with LoRA params $\Delta\theta$), pretrained filtering model $f_{\psi}$ trained on the same user interactions, historical sequences $\{S_u^{\le T}\}_{u\in\mathcal{U}}$, new full sequences $\{S_u^{A}\}_{u\in\mathcal{U}_A}$, threshold $t$, bottom-$K$, training epochs $E_f$ and $E_u$, learning rates $\eta$;
\ENSURE Updated LoRA parameters $\Delta\theta'$ on selected layers $\Phi$;
\STATE \textbf{Locate phase: identify globally sensitive layers}; 
\STATE Initialize selection counter $C(\ell) \leftarrow 0$ for each layer $\ell$;
\FOR{$u \in \mathcal{U}_A$}
    \STATE Compute embeddings $H^{\le T}_0 = E(I(S_u^{\le T})),\; H^{*}_0 = E(I(S_u^{A}))$;
    \FOR{layer $i = 1 \ldots L$}
        \STATE Forward to get $H^{\le T}_{\ell_i}, H^{A}_{\ell_i}$;
        \STATE $s_{\ell_i} \leftarrow \mathrm{cosine}(H^{A}_{\ell_i}, H^{\le T}_{\ell_i})$;
    \ENDFOR
    \STATE $\ell_{\text{sensitive}} \leftarrow \arg\min_i s_{\ell_i}$;
    \STATE $C(\ell_{\text{sensitive}}) \leftarrow C(\ell_{\text{sensitive}}) + 1$;
\ENDFOR
\STATE $\Phi \leftarrow$ Top-$t\%$ layers by $C(\ell)$;
\STATE \textbf{Forget phase: filtering outdated interactions};
\STATE Fine-tune $f_{\psi}$ on $\{S_u^{A}\}_{u\in\mathcal{U}_A}$ for $E_f$ epochs;
\FOR{$u \in \mathcal{U}_A$}
    \STATE Compute sequence embedding via $f_{\psi}$;
    \STATE Score each past interaction $S_u^{A-}$ via Eq.~(13);
    \STATE Remove bottom-$K$ items to get $(S_u^{A-})_{\text{filtered}}$;
    \STATE $S'_u \leftarrow (S_u^{A-})_{\text{filtered}} \oplus S_u^{A+}$;
\ENDFOR
\STATE Build instruction-tuning dataset $D'_A = \{ I(S'_u) \,|\, u \in \mathcal{U}_A \}$;
\STATE \textbf{Update phase: incremental fine-tuning};
\STATE Freeze all except LoRA params on layers $\Phi$;
\FOR{epoch $= 1 \ldots E_u$}
    \FOR{minibatch from $D'_A$}
        \STATE Compute $L_e$ on $u\in\mathcal{U}_A$ (Eq.~14);
        \STATE Compute $L_c$ on $u\in\mathcal{U}\setminus\mathcal{U}_A$ (Eq.~15);
        \STATE $L_{\text{total}} \leftarrow L_e + \lambda L_c$;
        \STATE Update $\Delta\theta|_{\Phi} \leftarrow \Delta\theta|_{\Phi} - \eta \nabla_{\Delta\theta|_{\Phi}} L_{\text{total}}$;
    \ENDFOR
\ENDFOR
\RETURN $\Delta\theta'$
\end{algorithmic}
\end{algorithm}
\section{Experimental Settings}
In this section, we present the data processing and experimental setup, aiming to address the following research questions:
\begin{itemize}[leftmargin=15pt] 
    \item \textbf{RQ1:} How does EvoRec improve the performance of $\mathcal{U}_A$ during preference updates and maintain stability under continuous updates, compared to other incremental learning methods, while preventing user preference forgetting?
    \item \textbf{RQ2:} How do different filtering models and the parameter localization strategy respectively contribute to the overall performance of EvoRec?
    \item \textbf{RQ3:} How do the choice of $t$, the value of $\lambda$, and the use of different large language models (e.g., LLaMA) influence the update performance of EvoRec?
    \item \textbf{RQ4:} How effective is the intuitive visualization of parameter localization?
    \item \textbf{RQ5:} How does EvoRec compare to other incremental learning methods in terms of efficiency in evolving recommendation models?
\end{itemize}

\label{Settings}
\subsection{\textbf{Datasets}}
\label{Settings:Datasets}
We conduct experiments on two real-world datasets, where the textual information in each dataset is standardized to use only product titles. The statistics of the processed data set are shown in Table \ref{tab:dataset_stats}, and its detailed descriptions are provided below. 
\begin{itemize}[leftmargin=15pt]
    \item \textbf{Amazon Beauty} is a subset of the Amazon review dataset\footnote{https://jmcauley.ucsd.edu/data/amazon/.}, specifically focused on user interactions with Beauty products. This dataset follows the 5-core setting, ensuring that each user and each item have received a minimum of five reviews. 
    \item \textbf{Amazon Toys} is another representative subset derived from the Amazon review dataset, containing user reviews of Toys products. Using the 5-core criterion, each user and each item in this data set have at least five associated reviews, providing a more comprehensive interaction history. 
\end{itemize}

\begin{table}[H]
\centering
\caption{Statistics of Datasets.}
\label{tab:dataset_stats}
\scalebox{0.9}{
\setlength{\tabcolsep}{14pt}
\begin{tabular}{@{}l|ccc}
\toprule
\textbf{Dataset} & \textbf{Beauty} & \textbf{Toys}  \\ \midrule
\# Users     & 15,139                & 11,822               \\
\# Items         & 12,094             & 11,864                 \\
\# Interaction before update & 130,457            & 98,298             \\
\# First update user count & 707            & 602  \\
\# Second update user count & 706            & 602  \\
\# Third update user count & 726            & 618  \\
\# Interaction after update & 141,645            & 106,480             \\ \bottomrule
\end{tabular}
}
\end{table}
We select three equally spaced timestamps from each dataset and divide all users into active and inactive groups according to whether they have new interactions before and after each timestamp. This means that three complete evolution cycles are performed on each dataset. The goal of EvoRec is to, across multiple evolution cycles, rapidly adapt to the latest preferences of active users while preserving the preferences of inactive users without forgetting. Each evolution is regarded as a complete cycle, during which we organized the original dataset in sequential form based on chronological order. At each timestamp $T$, users are divided into two groups, $\mathcal{U}_A$ and $\mathcal{U}_I$, depending on whether they have new interactions after $T$. Each sequential dataset is further split using the leave-one-out method, where the last interaction is used for testing, the second-to-last for validation, and the remaining interactions for training. To better evaluate performance in this evolutionary setting, results are reported separately for users in $\mathcal{U}_A$ and in the overall set $\{\mathcal{U}_A \cup \mathcal{U}_I\}$.

\subsection{\textbf{Baselines.}}
\label{Settings:Baselines}
In this section, we provide a detailed introduction to the seven backbone models utilized in our study: GRU4Rec~\cite{GRU4Rec}, SASRec~\cite{SASRec}, FMLP~\cite{fmlp}, TALLRec~\cite{bao2023tallrec}, TRSR~\cite{zheng2024harnessing}, LORAMoE~\cite{LoRAMoE}, and iLora~\cite{ilora}. Additionally, we present an in-depth discussion of six incremental learning paradigms: Re-training~\cite{sml}, Fine-tuning~\cite{hu2021lora}, FT-KL~\cite{FT-KL} ,EWC~\cite{ewc}, SML~\cite{sml}, and LSAT~\cite{lsta}.
The backbone models are described in detail as follows:
\begin{itemize}[leftmargin=15pt]
    \item \textbf{GRU4Rec}~\cite{GRU4Rec} is a method that uses a recurrent neural network (RNN) to model user sequences.
    \item \textbf{SASRec}~\cite{SASRec} is the first to apply attention mechanism to capture user sequential behaviors, enhancing modeling precision.
    \item \textbf{FMLP}~\cite{fmlp} uses filtering algorithms to reduce noise in logged user behavior data, improving the precision of sequential recommendation models.
    \item \textbf{TALLRec}~\cite{bao2023tallrec} learns to recommend based on prompts consisting solely of text and fine-tunes the LLMs using LoRA.
    \item \textbf{TRSR}~\cite{zheng2024harnessing} is a text-rich sequential recommendation framework that summarizes user history into compact prompts for efficient LLM-based recommendation, using hierarchical/recurrent summarization and LoRA fine-tuning.
    \item  \textbf{LORAMOE}~\cite{LoRAMoE} is a parameter-efficient fine-tuning method that combines multiple low-rank adapters (LoRA) with a router network in a Mixture-of-Experts (MoE) style. We apply this structure in TALLRec to mitigate forgetting of user interests on $\mathcal{U}_I$ during incremental updates.
    \item \textbf{iLoRA}~\cite{ilora} enhances sequential recommendation by integrating LoRA with a Mixture of Experts framework, dynamically adjusting parameters per user sequence via a gated expert selection mechanism to better capture individual preferences and reduce negative transfer.
\end{itemize}
The six incremental learning paradigms are described as follows:
\begin{itemize}[leftmargin=15pt]
      \item \textbf{Re-training}~\cite{sml} involves training a new model from scratch using the entire dataset, including the newly added data.
      \item \textbf{Fine-tuning}~\cite{hu2021lora} updates the existing model by training it solely on the newly added data.
       \item \textbf{FT-KL}~\cite{FT-KL}  incorporates KL divergence loss on top of continuous fine-tuning to constrain the model from deviating significantly from the original model.
      \item \textbf{EWC}~\cite{ewc} incorporates the EWC (Elastic Weight Consolidation) loss into the original model to mitigate forgetting in recommendation tasks.
    \item \textbf{SML}~\cite{sml} is a novel training paradigm designed to enhance the efficiency and accuracy of model retraining in recommender systems by leveraging past training experiences to adapt the old model to new data effectively.
    \item \textbf{LSAT}~\cite{lsta} is a framework designed to improve the performance of LLMRec by enabling incremental learning that captures both long-term and short-term user preferences.
\end{itemize}
For the backbone of traditional models (GRU4Rec~\cite{GRU4Rec}, SASRec~\cite{SASRec}, FMLP~\cite{fmlp}), we present five states: without updates, Re-training~\cite{sml}, Fine-tuning~\cite{hu2021lora}, EWC~\cite{ewc} and SML~\cite{sml}. For the backbone based on large language models (TALLRec~\cite{bao2023tallrec}, TRSR~\cite{zheng2024harnessing}, LORAMoE~\cite{LoRAMoE}, iLora~\cite{ilora}), we present seven states: without updates, Re-training~\cite{sml}, Fine-tuning~\cite{hu2021lora}, FT-KL~\cite{FT-KL}, EWC~\cite{ewc}, LSAT~\cite{lsta}, and our proposed EvoRec.
\subsection{\textbf{Implementation Details}}
\label{Settings:Implementation}
To ensure a fair comparison, we standardize the hyperparameters for traditional recommendation methods (GRU4Rec, SASRec, and FMLP) by setting the learning rate to 0.001, using the Adam optimizer, and configuring the batch size to 256 and the embedding dimension to 64. For TALLRec and TRSR, we employ Qwen2-7B-instruct\footnote{\url{https://huggingface.co/Qwen/Qwen2-7B-Instruct}} as the backbone model, set the learning rate to 0.0003, and utilize a LoRA adapter with a rank of 16. The initial model is trained for five epochs on an NVIDIA RTX 4090 (24GB). For LoRAMoE, we attach six LoRA experts and a gating network module to each FNN layer to enhance the modeling of user interest preferences and prevent interest forgetting. For iLoRA, we set the number of experts to four according to the original paper. All other settings are kept consistent with TALLRec and TRSR.

For incremental update strategies, the Re-training operation trains the model from scratch using the full dataset augmented with newly acquired interaction data. The fine-tuning operation updates the previous model using only the newly added interactions. When applying the EWC paradigm, we introduce a regularization term during fine-tuning to mitigate catastrophic forgetting by preventing excessive deviation from the original parameters, while FT-KL replaces the EWC loss with a KL divergence term and keeps all other settings the same. For SML, we follow the original paper’s setup to optimize the transfer parameters for each traditional backbone model. For LSAT, we train separate LoRA adapters for long-term and short-term user interests and determine the optimal fusion coefficient on the validation set for incremental learning.

For all traditional models, we generate the final recommendation list based on their predicted probability scores. For LLM-based recommendation methods (TALLRec, TRSR, LoRAMoE, and iLora ), we leverage the VLLM framework\footnote{\url{https://github.com/vllm-project/vllm}} for batch-accelerated inference to generate the recommended items.
\subsection{\textbf{Evaluation Metrics}}
\label{Settings:Evaluation}
We follow the setup of previous LLMRec studies~\cite{liao2024llara,ilora}. To ensure experimental robustness and fair comparison, we enlarge the candidate set size to 30 items for each prediction. Specifically, for every user interaction, we construct the candidate set by randomly sampling 29 non-interacted items as negative examples and including the ground-truth next item as the positive target. In this setup, LLMRec directly selects the most likely item from the candidate set, whereas traditional recommendation models adopt a ranking-based paradigm, where all items in the candidate set are first assigned predicted scores and then sorted accordingly.

For evaluation, we employ commonly used metrics in sequential recommendation, including Hit Rate (HR@1, HR@3) and Normalized Discounted Cumulative Gain (NDCG@3). HR measures whether the ground-truth item is correctly ranked within the top-$k$ positions, while NDCG takes into account the relative ranking positions and assigns higher importance to correctly ranking the target item near the top of the list. To provide a fair comparison across methods, we report results based on the top-5 predicted items with the highest probabilities or scores as the final recommendation outputs. For clarity, the best performance achieved by each method under every metric is highlighted in bold, which facilitates straightforward comparison between EvoRec and baseline approaches.
\section{Experiments}
\subsection{Overall Performance (RQ1)}

\begin{table}[!t]
\setlength{\tabcolsep}{4.2pt}
\centering
    \caption{Experimental Results on the Amazon Beauty. The first row under each base model (e.g., GRU4Rec) corresponds to its performance without any preference update, and the subsequent rows report the results after applying various update strategies. The best performance in each column is highlighted in bold and is statistically significant according to paired t-tests (p < 0.05).}
    \label{tab:performance_comparison_beauty}
\adjustbox{width=0.75\textwidth,height=\textheight,keepaspectratio}{
\begin{tabular}{l|l|cccccc}
    \toprule
    \multirow{3}{*}{\textbf{Backbone}} & \multirow{3}{*}{\textbf{Method}} & \multicolumn{6}{c}{\textbf{Amazon Beauty}} \\
\cmidrule(lr){3-8}
    &  & \multicolumn{3}{c}{Set of $\mathcal{U}_A$} & \multicolumn{3}{c}{Set of $\mathcal{U}$} \\
\cmidrule(lr){3-5} \cmidrule(lr){6-8}
    &  & HR@1 & HR@3 & NDCG@3 & HR@1 & HR@3 & NDCG@3 \\
\midrule
\multirow{5}{*}{GRU4Rec} & Base & 0.2312 & 0.3483 & 0.3043 & 0.2155 &0.3514 & 0.2971 \\
    & Re-training& 0.2595& 0.3583 & 0.3185 & 0.2197 & 0.3576 & 0.3003 \\
    & Fine-tuning & 0.2614 & 0.3642 & 0.3223 & 0.2105 & 0.3486 & 0.2917 \\
    & EWC & 0.2308 & 0.3497 & 0.3055 & 0.2162 & 0.3519 & 0.2991 \\
    & \textbf{SML} & \textbf{0.2638} & \textbf{0.3694} & \textbf{0.3289} &  \textbf{0.2214}& \textbf{0.3602} & \textbf{0.3114} \\
\midrule
\multirow{5}{*}{SASRec} & Base & 0.2659 & 0.3890 & 0.3363 & 0.2360 & 0.3769 & 0.3178 \\
    & Re-training & 0.2829 & 0.4031 & 0.3521 & 0.2410 & 0.3897 & 0.3271 \\
    & Fine-tuning & 0.2857 & 0.4097 & 0.3866 & 0.2259 & 0.3760 & 0.3154 \\
    & EWC & 0.2664 & 0.3905 & 0.3357 & 0.2362 & 0.3775 & 0.3154 \\
    & \textbf{SML} & \textbf{0.2928} & \textbf{0.4254} & \textbf{0.3902} & \textbf{0.2439} & \textbf{0.3884} & \textbf{0.3310} \\
\midrule
\multirow{5}{*}{FMLP} & Base & 0.2744 & 0.3974 & 0.3419 & 0.2497 & 0.3972 & 0.3350 \\
    & Re-training & 0.2841 & 0.4298 & 0.3679 & 0.2563 & \textbf{0.4094} & 0.3448 \\
    & Fine-tuning & 0.2926 & 0.4383 & 0.3795 & 0.2430 & 0.3953 & 0.3332 \\
    & EWC & 0.2789 & 0.3907 & 0.3436 & 0.2502 & 0.3987 & 0.3361 \\
    & \textbf{SML} & \textbf{0.3014} & \textbf{0.4485} & \textbf{0.3839} & \textbf{0.2621} & 0.4075 & \textbf{0.3498} \\
\midrule
\multirow{7}{*}{TALLRec} & Base & 0.4271 & 0.4526 & 0.4411 & 0.3607 & 0.4046 & 0.3852 \\
    & Re-training & 0.4379 & 0.4649 & 0.4523 & 0.3721 & 0.4139 & 0.3965 \\
    & Fine-tuning & 0.4413 & 0.4653 & 0.4548 & 0.3559 & 0.3936 & 0.3781 \\
    & FT-KL & 0.4385 & 0.4621 & 0.4512 & 0.3592 & 0.3974 & 0.3819 \\
    & WEC & 0.4371 & 0.4609 & 0.4498 & 0.3698 & 0.4122 & 0.3942 \\
    & LSAT & 0.4522 & 0.4789 & 0.4551 & 0.3624 & 0.4085 & 0.3847 \\
    & \textbf{EvoRec} & \textbf{0.4601} &\textbf{0.4912} &\textbf{0.4762} & \textbf{0.3776} &\textbf{0.4157} & \textbf{0.3991} \\
\midrule
\multirow{7}{*}{TRSR} & Base & 0.4032 & 0.4329 & 0.4195 & 0.3332 & 0.3795 & 0.3587 \\
    & Re-training & 0.4265 & 0.4542 & 0.4374 & 0.3420 & 0.3861 & 0.3671 \\
    & Fine-tuning & 0.4185 & 0.4492 & 0.4241 & 0.3156 & 0.3534 & 0.3329 \\
    & FT-KL & 0.4156 & 0.4461 & 0.4207 & 0.3189 & 0.3572 & 0.3367 \\
    & WEC & 0.4142 & 0.4449 & 0.4194 & 0.3398 & 0.3844 & 0.3654 \\
    & LSAT & 0.4326 & 0.4679 & 0.4439 & 0.3215 & 0.3729 & 0.3516 \\
    & \textbf{EvoRec} & \textbf{0.4463} &\textbf{0.4759} &\textbf{0.4531} & \textbf{0.3425} &\textbf{0.3879} & \textbf{0.3691} \\
\midrule
\multirow{7}{*}{LoRAMoE} & Base & 0.4357& 0.4617& 0.4482& 0.3694& 0.4126& 0.3932 \\
    & Re-training & 0.4526 &	0.4783&	0.4629&	0.3731&	0.4158&	0.3975 \\
    & Fine-tuning & 0.4462 & 0.4725&	0.4562&	0.3419&	0.3865&	0.3711 \\
    & FT-KL & 0.4434 & 0.4693 & 0.4527 & 0.3452 & 0.3903 & 0.3749 \\
    & WEC & 0.4421 & 0.4681 & 0.4514 & 0.3708 & 0.4141 & 0.3958 \\
    & LSAT & 0.4503&0.4771&	0.4614&	0.3632&	0.4065&	0.3854 \\
    & \textbf{EvoRec} & \textbf{0.4671} &\textbf{0.4939} &\textbf{0.4762} & \textbf{0.3751} &\textbf{0.4169} & \textbf{0.3992} \\
\midrule
\multirow{7}{*}{iLora} & Base & 0.4521 & 0.4776 & 0.4661 & 0.3857 & 0.4296 & 0.4102 \\
    & Re-training & 0.4589 & 0.4847 & 0.4731 & 0.3906 & \textbf{0.4317} & 0.4127 \\
    & Fine-tuning & 0.4602 & 0.4862 & 0.4745 & 0.3705 & 0.4201 & 0.4012 \\
    & FT-KL & 0.4530 & 0.4785 & 0.4670 & 0.3869 & 0.4309 & 0.4115 \\
    & WEC & 0.4526 & 0.4781 & 0.4666 & 0.3874 & 0.4313 & 0.4118 \\
    & LSAT & 0.4602 & 0.4862 & 0.4745 & 0.3741 & 0.4245 & 0.4026 \\
    & \textbf{EvoRec} & \textbf{0.4821} &\textbf{0.5076} &\textbf{0.4961} & \textbf{0.3954} &\textbf{0.4348} & \textbf{0.4159} \\
    \bottomrule
\end{tabular}
}
\end{table}

\begin{table}[!t]
\setlength{\tabcolsep}{4.2pt}
\centering
\caption{Experimental Results on the Amazon Toys. The first row under each base model (e.g., GRU4Rec) corresponds to its performance without any preference update, and the subsequent rows report the results after applying various update strategies. The best performance in each column is highlighted in bold and is statistically significant according to paired t-tests (p < 0.05).}
\label{tab:performance_comparison_toys}
\adjustbox{width=0.75\textwidth,height=\textheight,keepaspectratio}{
\begin{tabular}{l|l|cccccc}
\toprule
\multirow{3}{*}{\textbf{Backbone}} & \multirow{3}{*}{\textbf{Method}} & \multicolumn{6}{c}{\textbf{Amazon Toys}} \\
\cmidrule(lr){3-8}
    &  & \multicolumn{3}{c}{Set of $\mathcal{U}_A$} & \multicolumn{3}{c}{Set of $\mathcal{U}$} \\
\cmidrule(lr){3-5} \cmidrule(lr){6-8}
    &  & HR@1 & HR@3 & NDCG@3 & HR@1 & HR@3 & NDCG@3 \\
\midrule
\multirow{5}{*}{GRU4Rec} & Base & 0.1134 & 0.2346 & 0.1891 & 0.1654 & 0.3027 & 0.2631 \\
    & Re-training & 0.1289 & 0.2504 & 0.1943 & 0.1701 & 0.3105 & 0.2644 \\
    & Fine-tuning & 0.1358 & 0.2609 & 0.1988 & 0.1698 & 0.3105 & 0.2599 \\
    & EWC & 0.1150 & 0.2371 & 0.1908 & 0.1662 & 0.3045 & 0.2649 \\
    & \textbf{SML} & \textbf{0.1370} & \textbf{0.2652} &\textbf{0.2017} & \textbf{0.1723} & \textbf{0.3202} & \textbf{0.2682} \\
\midrule
\multirow{5}{*}{SASRec} & Base & 0.1412 & 0.2774 & 0.2193 & 0.1977 & 0.3432 & 0.2819 \\
    & Re-training & 0.1604 & 0.2927 & 0.2233 & 0.2102 & 0.3574 & 0.2933 \\
    & Fine-tuning & 0.1728 & 0.3272 & 0.2613 & 0.1853 & 0.3244 & 0.2623 \\
    & EWC & 0.1424 & 0.2785 & 0.2190 & 0.1980 & 0.3439 & 0.2821 \\
    & \textbf{SML} & \textbf{0.1735} & \textbf{0.3310} & \textbf{0.2639} & \textbf{0.2118} & \textbf{0.3586} & \textbf{0.2947} \\
\midrule
\multirow{5}{*}{FMLP} & Base & 0.1561 & 0.2608 & 0.2164 & 0.2259 & 0.3573 & 0.3054 \\
    & Re-training & 0.1727 & 0.2841 & 0.2232 & 0.2278 & \textbf{0.3642} & 0.3125 \\
    & Fine-tuning & 0.1844 & 0.3123 & 0.2596 & 0.2063 & 0.3321 & 0.2908 \\
    & EWC & 0.1570 & 0.2613 & 0.2175 & 0.2262 & 0.3579 & 0.3056 \\
    & \textbf{SML} & \textbf{0.1742} & \textbf{0.2915} & \textbf{0.2619} & \textbf{0.2302} & 0.3589 & \textbf{0.3171} \\
\midrule
\multirow{7}{*}{TALLRec} & Base & 0.3387 & 0.3737 & 0.3582 & 0.3208 & 0.3676 & 0.3475 \\
    & Re-training & 0.3674 & 0.4056 & 0.3948 & 0.3285 & 0.3749 & 0.3580 \\
    & Fine-tuning & 0.3338 & 0.3837 & 0.3616 & 0.3055 & 0.3547 & 0.3335 \\
    & FT-KL & 0.3359 & 0.3804 & 0.3581 & 0.3088 & 0.3585 & 0.3373 \\
    & EWC & 0.3345 & 0.3792 & 0.3568 & 0.3262 & 0.3732 & 0.3563 \\
    & LSAT & 0.3538 & 0.3970 & 0.3734 & 0.3272 & 0.3788 & 0.3507 \\
    & \textbf{EvoRec} & \textbf{0.3872} & \textbf{0.4285} & \textbf{0.4089} & \textbf{0.3356} & \textbf{0.3813} & \textbf{0.3629} \\
\midrule
\multirow{7}{*}{TRSR} & Base & 0.3438 & 0.3853 & 0.3672 & 0.3215 & 0.3671 & 0.3474 \\
    & Re-training & 0.3619 & 0.4023 & 0.3821 & 0.3294 & 0.3731 & 0.3514 \\
    & Fine-tuning & 0.3239 & 0.3654 & 0.3472 & 0.3037 & 0.3505 & 0.3299 \\
    & FT-KL & 0.3289 & 0.3724 & 0.3542 & 0.3070 & 0.3543 & 0.3337 \\
    & EWC & 0.3275 & 0.3712 & 0.3529 & 0.3271 & 0.3714 & 0.3497 \\
    & LSAT & 0.3542 & 0.3961 & 0.3765 & 0.3195 & 0.3636 & 0.3445 \\
    & \textbf{EvoRec} & \textbf{0.3721} & \textbf{0.4117} & \textbf{0.3946} & \textbf{0.3318} & \textbf{0.3774} & \textbf{0.3576} \\
\midrule
\multirow{7}{*}{LoRAMoE} & Base & 0.3562& 0.3941& 0.3747& 0.3327& 0.3785& 0.3562 \\
    & Re-training & 0.3675&	0.4051&	0.3836&	0.3406&	0.3861&	0.3642 \\
    & Fine-tuning & 0.3621 & 0.4008&	0.3819&	0.3157&	0.3542&	0.3318 \\
    & FT-KL & 0.3593 & 0.3976 & 0.3784 & 0.3190 & 0.3580 & 0.3356 \\
    & EWC & 0.3579 & 0.3964 & 0.3771 & 0.3383 & 0.3844 & 0.3625 \\
    & LSAT & 0.3656&0.4028&	0.3846&	0.3284&	0.3744&	0.3529 \\
    & \textbf{EvoRec} & \textbf{0.3813} & \textbf{0.4195} & \textbf{0.3972} & \textbf{0.3471} & \textbf{0.3923} &\textbf{ 0.3715} \\
\midrule
\multirow{7}{*}{iLora} & Base & 0.3612 & 0.3981 & 0.3824 & 0.3453 & 0.3928 & 0.3718 \\
    & Re-training & 0.3665 & 0.4036 & 0.3879 & 0.3486 & 0.3957 & 0.3744 \\
    & Fine-tuning & 0.3676 & 0.4049 & 0.3892 & 0.3378 & 0.3851 & 0.3642 \\
    & FT-KL & 0.3625 & 0.3995 & 0.3838 & 0.3465 & 0.3941 & 0.3731 \\
    & EWC & 0.3620 & 0.3989 & 0.3832 & 0.3471 & 0.3947 & 0.3736 \\
    & LSAT & 0.3676 & 0.4049 & 0.3892 & 0.3410 & 0.3885 & 0.3675 \\
    & \textbf{EvoRec} & \textbf{0.3924} & \textbf{0.4291} & \textbf{0.4133} & \textbf{0.3487} & \textbf{0.3966} & \textbf{0.3752} \\
\bottomrule
\end{tabular}
}
\end{table}

\begin{figure}[H]
\setlength{\abovecaptionskip}{0.2cm}
  \centering

  \begin{subfigure}{0.4\textwidth}
    \includegraphics[width=\linewidth]{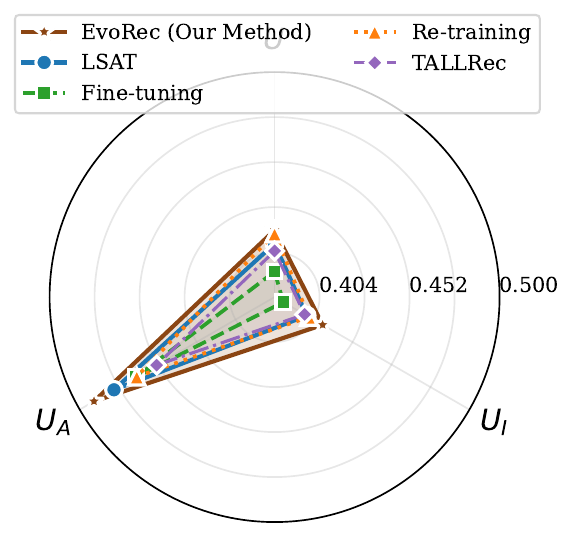}
    \caption{HR@3 on Beauty}
  \end{subfigure}
  \begin{subfigure}{0.4\textwidth}
    \includegraphics[width=\linewidth]{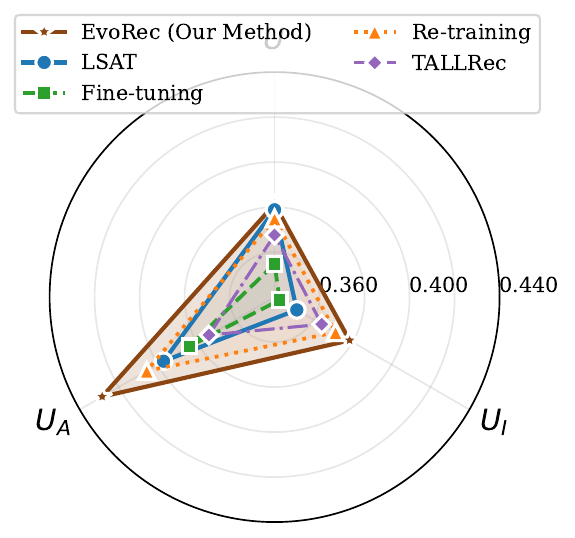}
    \caption{HR@3 on Toys}
  \end{subfigure}

  \caption{Radar chart illustrating the performance of different incremental learning paradigms on 
$\mathcal{U}_A$, $\mathcal{U}_I$, and $\mathcal{U}$ after a single update (using TALLRec as the recommendation method).}
  \label{fig:radar}
\end{figure}

We conduct a detailed comparison of the proposed EvoRec with other incremental learning paradigms at time $T$. The detailed results are presented in Table~\ref{tab:performance_comparison_beauty}, ~\ref{tab:performance_comparison_toys} and Figure~\ref{fig:radar}. Based on the experimental analysis, we draw the following conclusions:

For all backbone models, testing the unupdated models with the latest data reveals suboptimal performance on $\mathcal{U}_A$. This observation indicates that employing outdated models for users whose preferences have shifted is ineffective, highlighting the necessity of incremental learning. After performing Re-training, the updated models exhibit significant performance improvements over the unupdated models on both $\mathcal{U}_A$ and the overall user set $\mathcal{U}$, demonstrating that Re-training is a stable and effective incremental update method. However, training the model from scratch incurs the highest computational cost among all methods. Examining the fine-tuning results, we observe a notable performance improvement on $\mathcal{U}_A$, with some backbone models (e.g., TALLRec and FMLP) even surpassing the performance of  Re-training. This suggests that fine-tuning is particularly effective at capturing users’ short-term preferences. In contrast, fine-tuning exhibits the catastrophic forgetting effect inherent in incremental learning when evaluated on $\mathcal{U}$. Furthermore, experimental results indicate that EWC can effectively mitigate the forgetting of user preferences to some extent, although the performance on $\mathcal{U}_A$ does not show a significant improvement after applying it. The SML method achieves the best average performance on non-LLM-based methods, but due to the large parameter size of LLMs, it is not applicable to LLM-based recommendation methods (LLMRec).

As shown in Figure~\ref{fig:radar}, EvoRec consistently encloses the largest area across all three user sets, outperforming other incremental learning paradigms. It thus achieves the best performance on $\mathcal{U}_A$, $\mathcal{U}_I$, and $\mathcal{U}$. Compared to the expensive training cost of Re-training, EvoRec does not require training from scratch. Additionally, EvoRec outperforms fine-tuning and LSAT on $\mathcal{U}_A$, indicating that filtering out outdated interactions and modeling only the latest interactions better enables LLMs to capture users' short-term preferences. On $\mathcal{U}_I$, EvoRec does not suffer from performance degradation; instead, it achieves a slight performance gain. This result validates the effectiveness of our proposed old-new preference localization method, which accurately identifies the sensitive parameters responsible for storing user-specific knowledge. Updating only these parameters effectively preserves the model’s memory of other users' preferences, ensuring that incremental updates do not degrade overall recommendation performance.

\subsection{\textbf{In-Depth Analysis}}
\subsubsection{\textbf{Evaluating EvoRec Under Continuous Incremental Updates (RQ1)}}
To evaluate the effectiveness of EvoRec (using TALLRec as the recommendation method) under continuous incremental updates, we conduct three updates over three consecutive periods, with the experimental results presented in Figure~\ref{fig:continuous}. We aggregate all users updated during the three periods as $\mathcal{U}_A$, and we also report the performance changes of the overall user $\mathcal{U}$. The results indicate that, compared to other incremental learning methods, EvoRec consistently achieves the most significant performance improvement for users in $\mathcal{U}_A$ after each update. Observing the overall performance on $\mathcal{U}$, fine-tuning leads to a severe decline in performance due to user preference forgetting. LSAT introduces a LoRA adapter in each update to capture short-term user preferences, resulting in notable improvements in $\mathcal{U}_A$. However, the continuous introduction of additional parameters causes a slight performance drop in $\mathcal{U}$. In contrast, EvoRec maintains stable performance on $\mathcal{U}$ without significant preference forgetting, demonstrating its effectiveness for continuous incremental updates.
\begin{figure}[H]
\setlength{\abovecaptionskip}{0.2cm}
  \centering

  \begin{subfigure}{0.245\textwidth}
    \includegraphics[width=\linewidth]{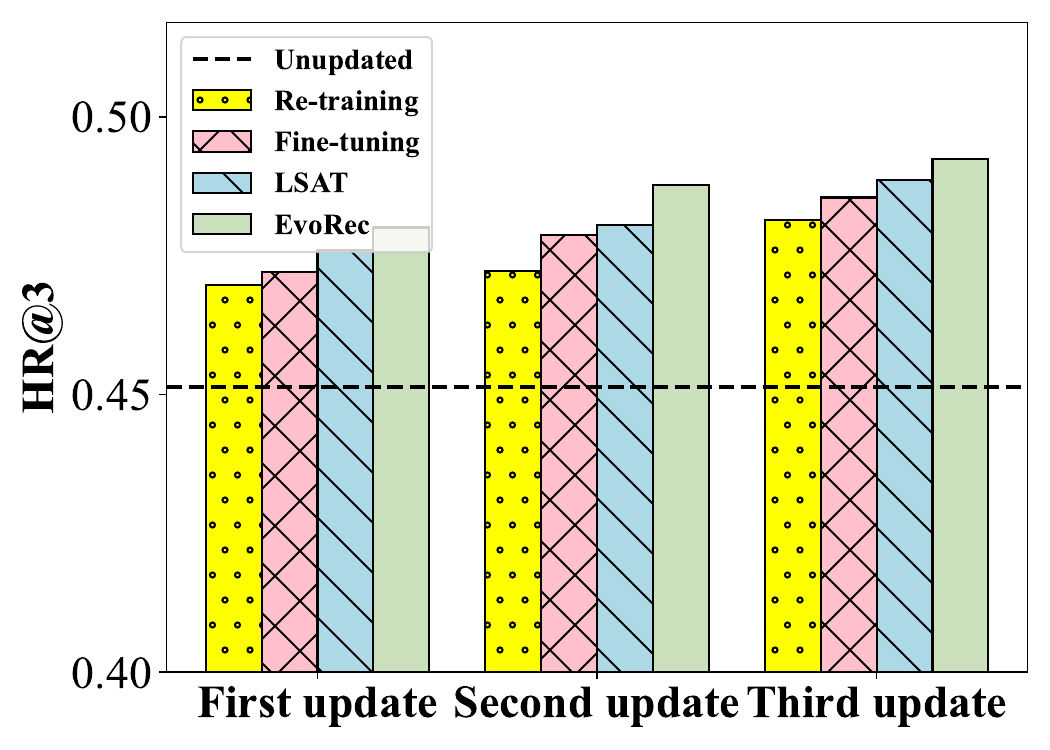}
    \caption{HR@3 on $\mathcal{U}_A$}
  \end{subfigure}
  \begin{subfigure}{0.245\textwidth}
    \includegraphics[width=\linewidth]{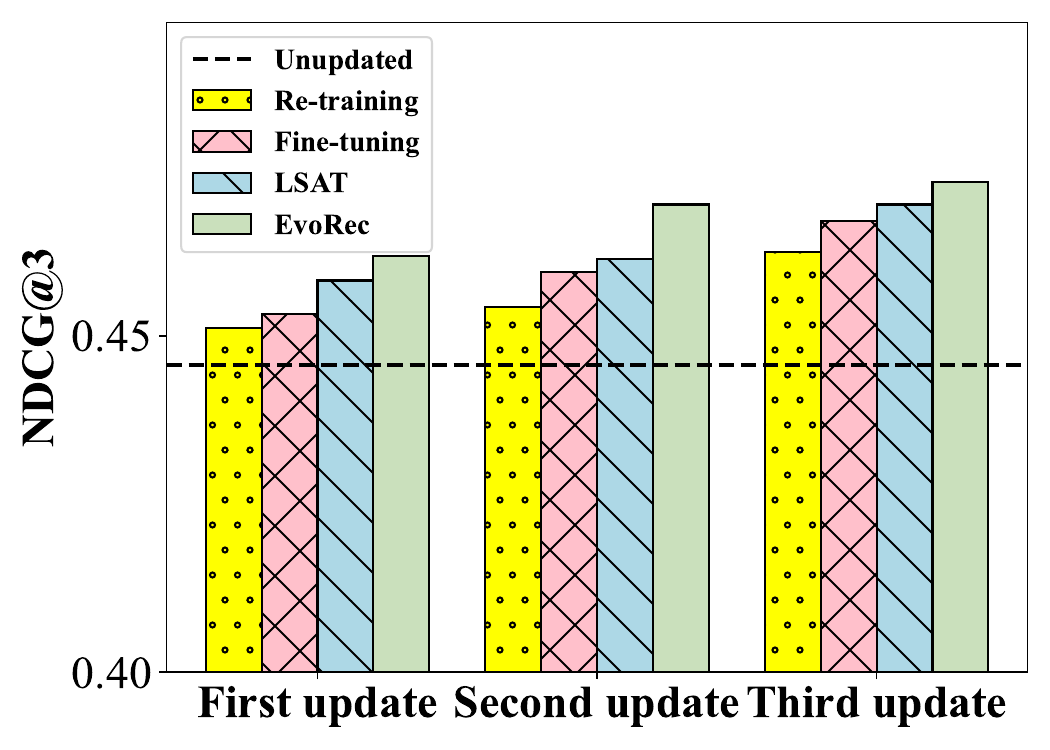}
    \caption{NDCG@3 on $\mathcal{U}_A$}
  \end{subfigure}
  \begin{subfigure}{0.245\textwidth}
    \includegraphics[width=\linewidth]{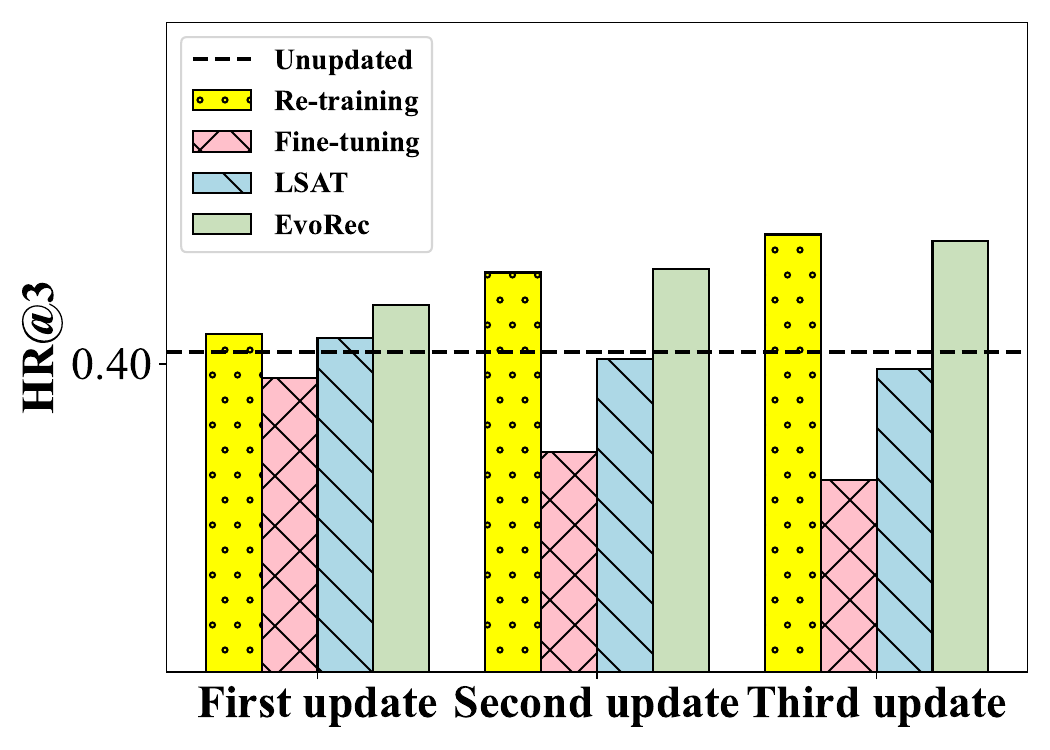}
    \caption{HR@3 on $\mathcal{U}$}
  \end{subfigure}
  \begin{subfigure}{0.245\textwidth}
    \includegraphics[width=\linewidth]{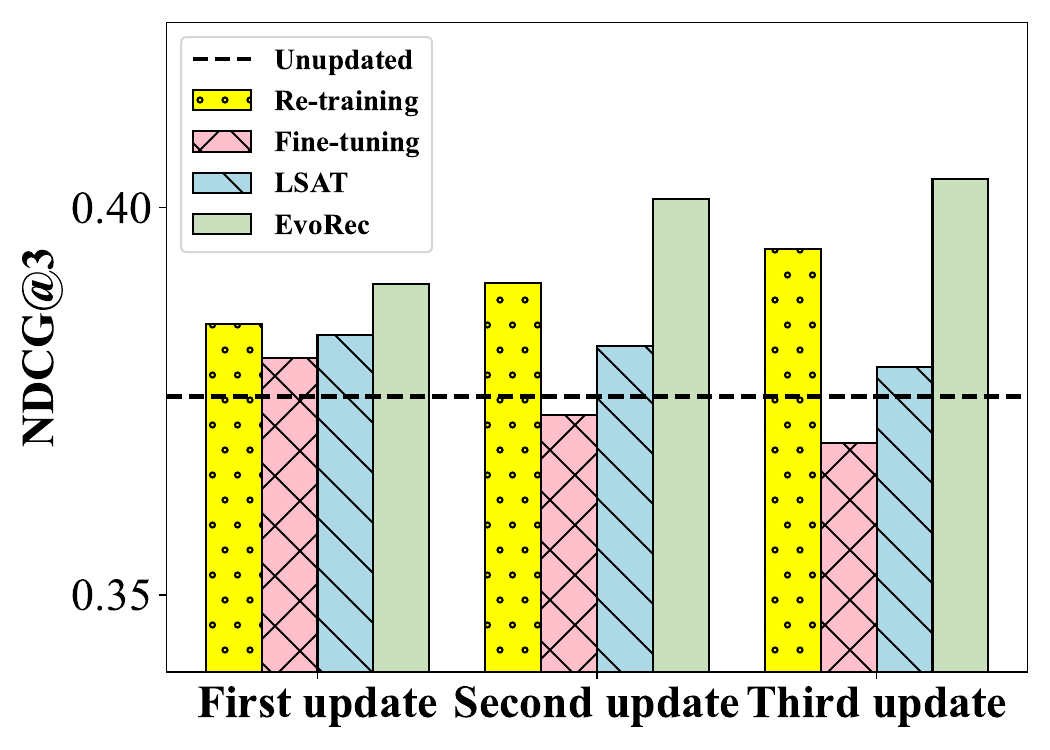}
    \caption{NDCG@3 on $\mathcal{U}$}
  \end{subfigure}
  \caption{The performance variations of EvoRec after three updates on Beauty.}
  \label{fig:continuous}
\end{figure}

\begin{figure}[H]
\setlength{\abovecaptionskip}{0.2cm}
  \centering
  \begin{subfigure}{0.245\textwidth}
    \includegraphics[width=\linewidth]{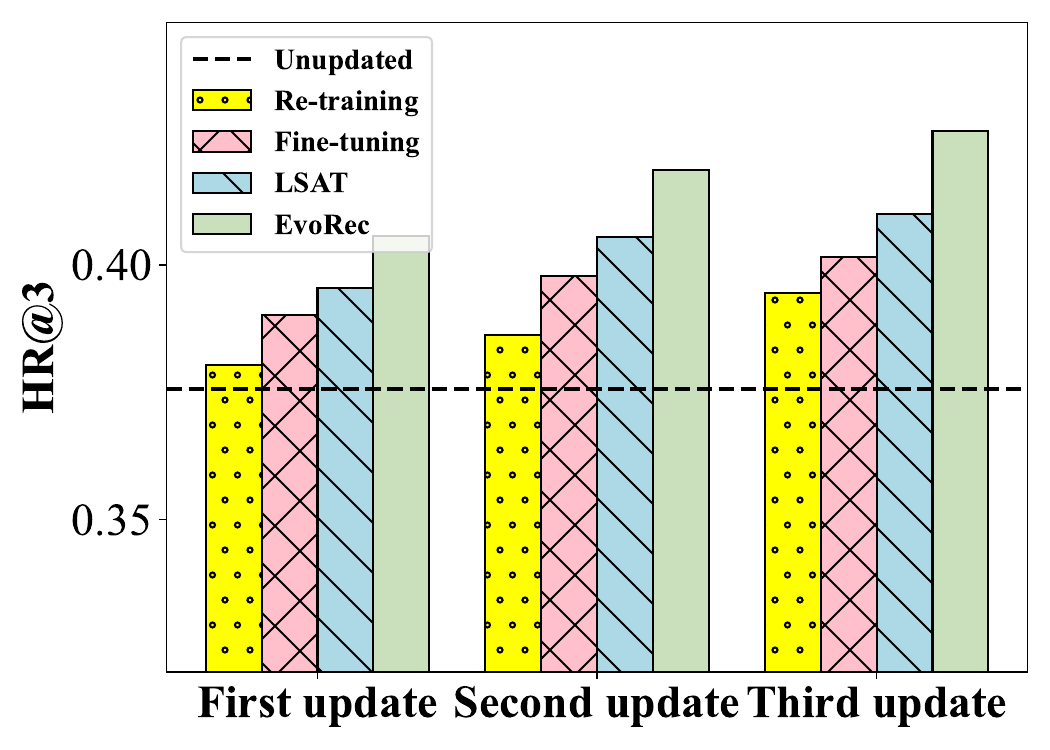}
    \caption{HR@3 on $\mathcal{U}_A$}
  \end{subfigure}
  \begin{subfigure}{0.245\textwidth}
    \includegraphics[width=\linewidth]{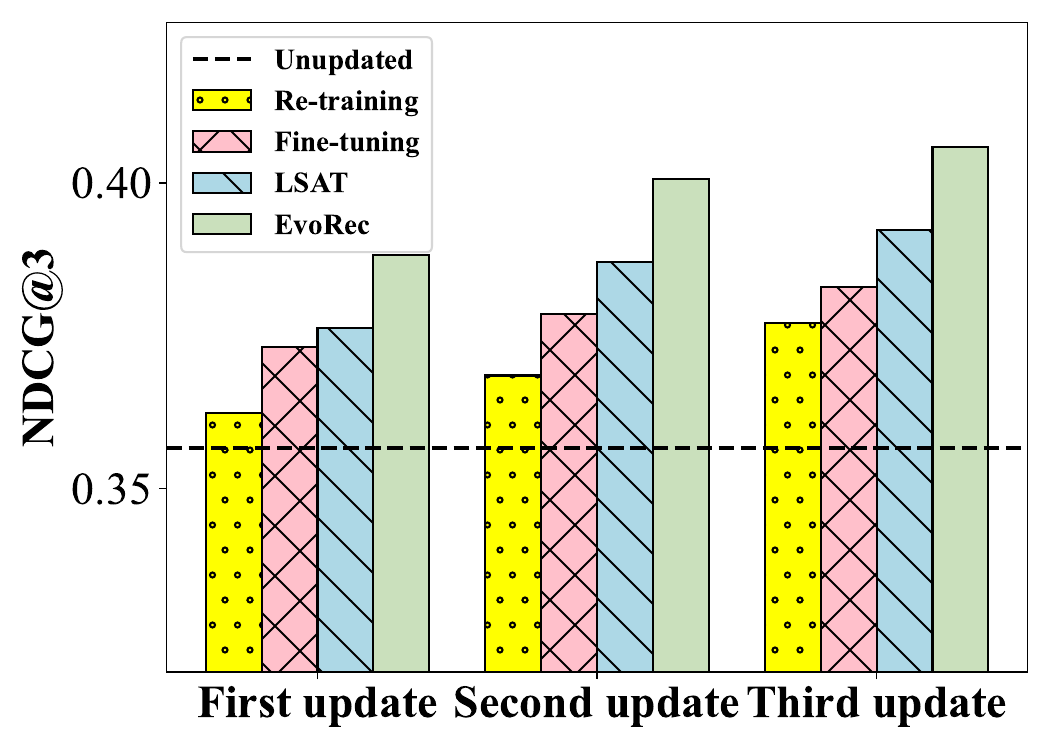}
    \caption{NDCG@3 on $\mathcal{U}_A$}
  \end{subfigure}
  \begin{subfigure}{0.245\textwidth}
    \includegraphics[width=\linewidth]{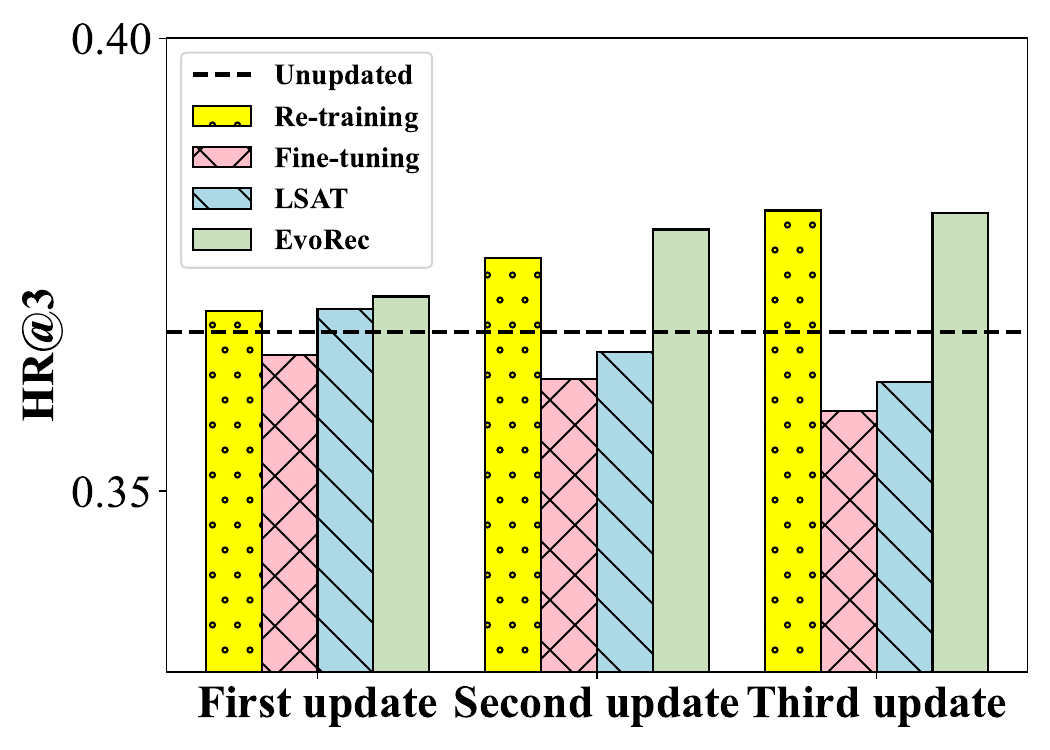}
    \caption{HR@3 on $\mathcal{U}$}
  \end{subfigure}
  \begin{subfigure}{0.245\textwidth}
    \includegraphics[width=\linewidth]{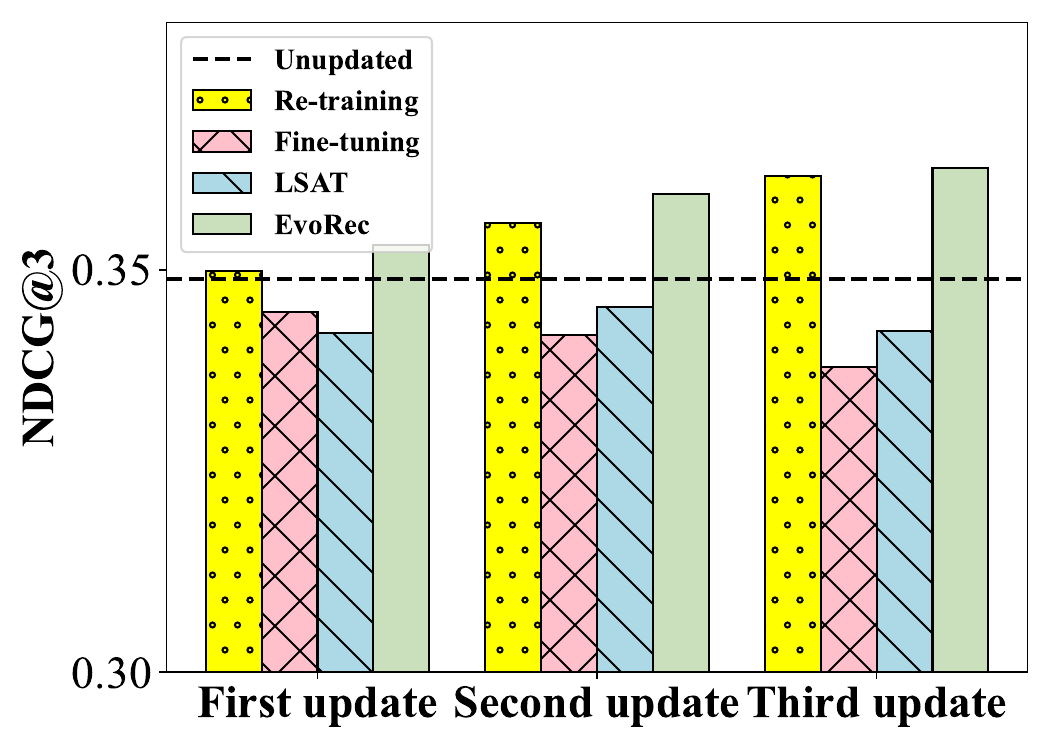}
    \caption{NDCG@3 on $\mathcal{U}$}
  \end{subfigure}

  \caption{The performance variations of EvoRec after three updates on Toys.}
  \label{fig:continuous}
\end{figure}
\subsubsection{\textbf{Ablation Study (RQ2)}}

\begin{table}[htbp]
\centering
\caption{Ablation study of EvoRec components on different backbones.}
\label{tab:Ablation}
\scalebox{0.95}{  
\begin{tabular}{l l|ccc|ccc}
\toprule
\multirow{2}{*}{\textbf{Backbone}} & \multirow{2}{*}{\textbf{Variants}} 
& \multicolumn{3}{c|}{\textbf{Amazon Beauty}} 
& \multicolumn{3}{c}{\textbf{Amazon Toys}} \\
\cmidrule(lr){3-5} \cmidrule(lr){6-8}
& & HR@1 & HR@3 & NDCG@3 & HR@1 & HR@3 & NDCG@3 \\
\midrule

\multirow{3}{*}{TALLRec}
& w/o Location         & 0.4389 & 0.4681 & 0.4571 & 0.3452 & 0.3837 & 0.3616 \\
& w/o Filtering Model  & 0.4509 & 0.4823 & 0.4639 & 0.3722 & 0.4137 & 0.3946 \\
& EvoRec               & 0.4601 & 0.4912 & 0.4762 & 0.3872 & 0.4285 & 0.4089 \\
\midrule

\multirow{3}{*}{TRSR}
& w/o Location         & 0.4121 & 0.4435 & 0.4263 & 0.3511 & 0.3920 & 0.3756 \\
& w/o Filtering Model  & 0.4333 & 0.4629 & 0.4418 & 0.3656 & 0.4029 & 0.3857 \\
& EvoRec               & 0.4463 & 0.4759 &  0.4531 & 0.3721 &  0.4117 & 0.3946 \\
\midrule

\multirow{3}{*}{LORAMOE}
& w/o Location         & 0.4431 & 0.4695 & 0.4525 & 0.3629 & 0.4011 & 0.3826 \\
& w/o Filtering Model  & 0.4588 & 0.4839 & 0.4644 & 0.3756 & 0.4062 & 0.3861 \\
& EvoRec               & 0.4671 & 0.4939 & 0.4762 & 0.3813 & 0.4195 &  0.3972 \\

\bottomrule
\end{tabular}
}
\end{table}


To investigate the importance of each component in EvoRec, we design two variant models based on the original EvoRec: w/o Location (randomly selecting the same number of parameters as the sensitive parameters) and w/o Filtering Model (removing the filtering effect of the filtering model). The experimental results are presented in Table~\ref{tab:Ablation} (using TALLRec, TRSR, and LORAMOE as the recommendation methods). On the set $\mathcal{U}_A$, the w/o Location variants of all three LLM-based recommendation methods consistently underperform EvoRec, indicating that the sensitive parameters located by EvoRec are the most critical ones for modeling preference shifts. Updating only these parameters not only enables rapid adaptation to users' latest interests but also prevents large-scale parameter modifications from causing preference forgetting for other users. The w/o Filtering Model variant shows a performance drop on $\mathcal{U}_A$, suggesting that some outdated interactions from the previous period have a negative impact on user preference modeling, and further indicating that LLMs are more adept at capturing short-term user interactions. Therefore, removing outdated interactions from the sequence is necessary. The results across all three backbone recommendation methods confirm that the filtering model in EvoRec can efficiently eliminate obsolete interactions, thereby improving the accuracy of sequential modeling.

\subsection{\textbf{Visualization of Localized Parameters (RQ4)}}
\label{analysis:Localized}
\begin{figure}[ht]
    \centering
     \includegraphics[width=1.0\textwidth]{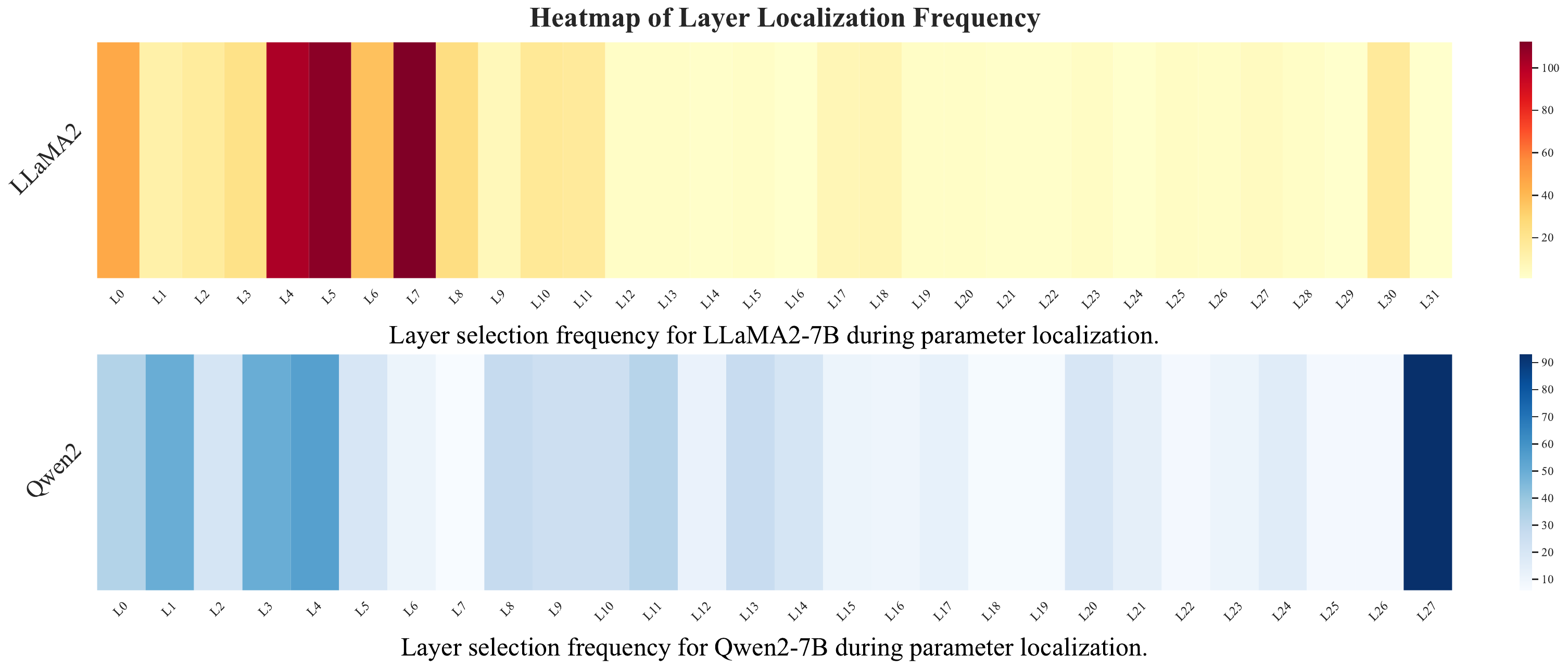}
    \caption{Heatmap of Sensitive Parameter Distribution on the Toys Dataset.}
    \label{fig:heatmap}
\end{figure}

 In this section, we analyze the sensitivity distribution across layers during the localization process. Specifically, we present the frequency with which each layer in Qwen2-7B and LLaMA2-7B is identified as sensitive on the Toys dataset, as shown in Figure~\ref{fig:heatmap}. The experimental results show that the identified layers are selected with significantly higher frequency than other layers, indicating that these parameters play a crucial role in the transition between users’ past and emerging preferences. Updating only these sensitive parameters enables efficient adaptation to evolving user interests while preserving recommendation performance for other users. Moreover, the most frequently selected (darker) layers are not necessarily concentrated in the deeper parts of the LLMs. Due to differences in backbone architectures, the localization results vary even on the same dataset. This demonstrates that EvoRec adaptively selects sensitive parameters based on the current model's understanding of user interests, rather than consistently targeting the same layers.

\subsection{Backbone Analysis (RQ3)}
\label{analysis:Backbone}
\begin{figure}[ht]
    \centering
     \includegraphics[width=1.0\textwidth]{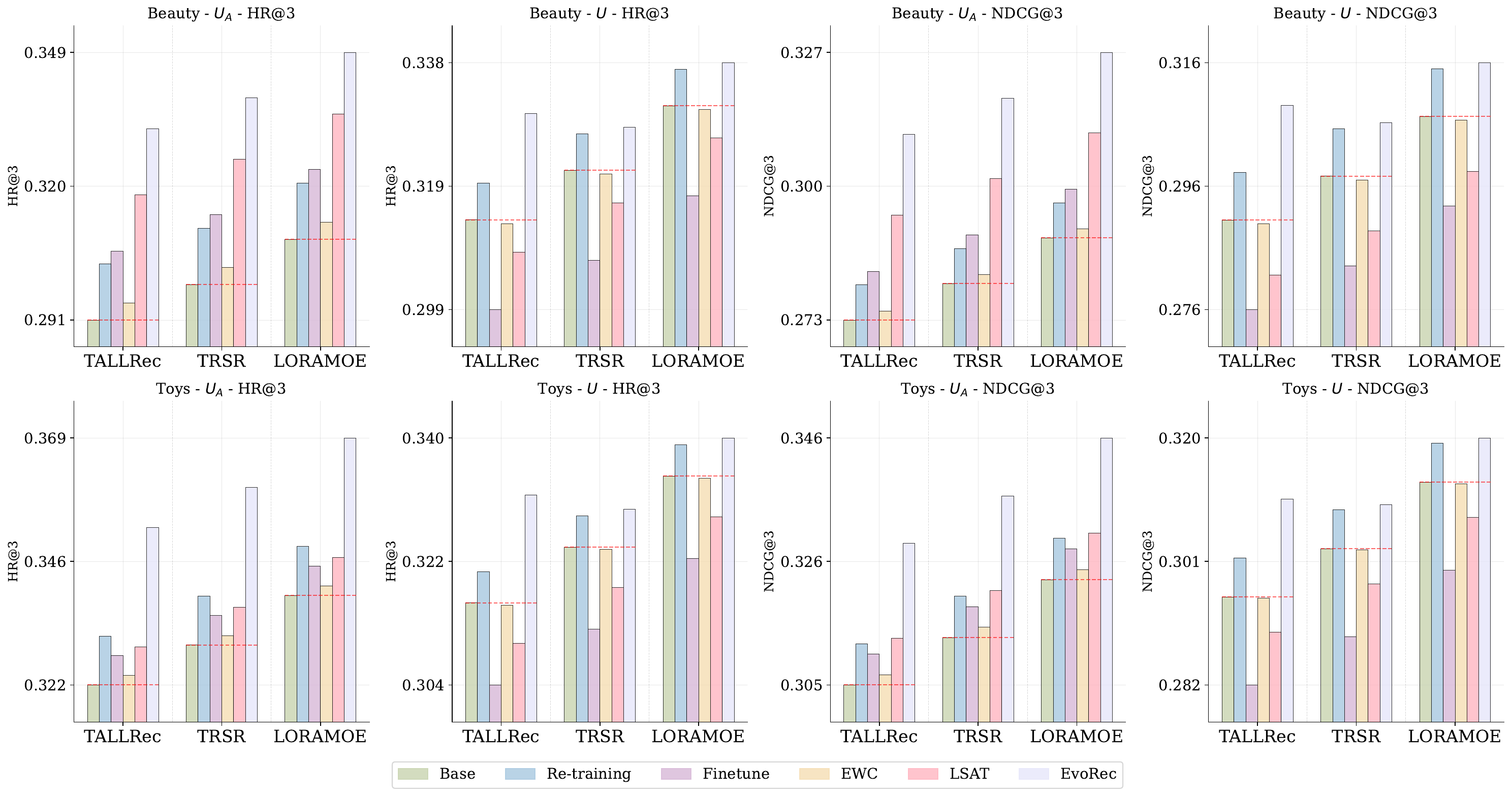}
    \caption{Performance of Different Evolution Algorithms on LLaMA-2-7B-Chat.}
    \label{fig:llama_as_bcakbone}
\end{figure}
We replace the backbone models of three LLM-based recommendation algorithms (TALLRec, TRSR, and LORAMOE) with the LLaMA-2-7B-Chat model\footnote{\url{https://huggingface.co/meta-llama/Llama-2-7b-chat-hf}}. The experimental results are shown in Figure~\ref{fig:llama_as_bcakbone}.

As illustrated in the figure, all three recommendation algorithms exhibit performance improvements on the user set $\mathcal{U}_A$, with EvoRec, LSAT, and Re-training showing the most notable gains, while EWC demonstrates the least improvement. On the full user set $\mathcal{U}$, most incremental learning paradigms, except EvoRec and Re-training, suffer from performance degradation due to user preference forgetting. Among them, Fine-tuning and LSAT experience the most significant drops in performance. This suggests that large-scale modifications of LoRA parameters, although beneficial for users in $\mathcal{U}_A$, lead to considerable overall performance degradation. EWC imposes constraints to prevent drastic parameter shifts before and after model evolution, which helps mitigate preference forgetting to some extent, but yields only limited improvement on $\mathcal{U}_A$. Re-training achieves strong performance on both $\mathcal{U}_A$ and $\mathcal{U}$, but at the cost of substantial time and computational resources. In contrast, EvoRec avoids user preference forgetting, efficiently adapts to the interest shifts in $\mathcal{U}_A$, and requires significantly less time and computation compared to Re-training. The results on the LLaMA-2-7B-Chat model demonstrate that EvoRec exhibits strong adaptability and robustness across different LLMs, enabling efficient model evolution.

\subsection{Analysis of the Filtering Model and Hyperparameter (RQ3)}
\label{analysis:Filtering}

   
\begin{figure}[ht]
    \centering
    \begin{subfigure}{1.0\textwidth}
        \centering
        \includegraphics[width=\textwidth]{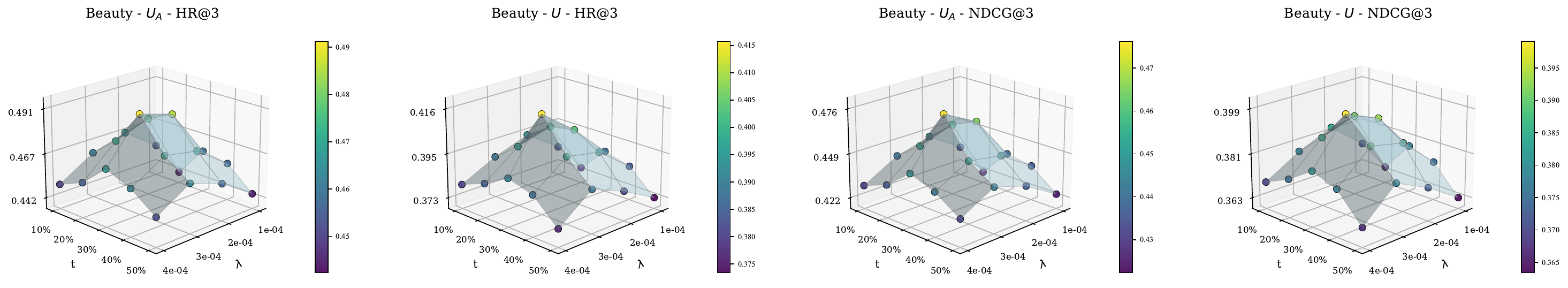}
        \label{fig:heatmap_beauty}
    \end{subfigure}
    \vspace{0.5em}  
    \begin{subfigure}{1.0\textwidth}
        \centering
        \includegraphics[width=\textwidth]{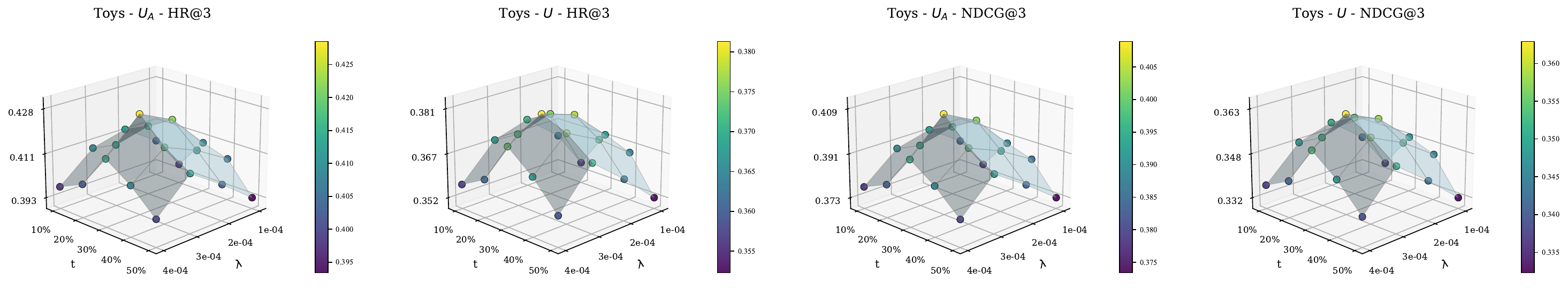}
        \label{fig:heatmap_toys}
    \end{subfigure}
    \caption{Parameter Sensitivity Analysis on Beauty and Toys.}
    \label{fig:3d}
\end{figure}

The hyperparameters $\lambda$ and $t$ are designed to control the weight of the KL divergence term and the threshold for selecting sensitive parameters in the EvoRec model, respectively. These parameters play a critical role in enhancing the model’s evolution capability while preventing user preference forgetting. Specifically, $\lambda$ regulates the contribution of the KL divergence loss in the overall objective, constraining the deviation of the evolved model from the original parameters, thereby facilitating stable updates without catastrophic forgetting.

As shown in Figure~\ref{fig:3d}, increasing the value of $\lambda$ leads to a performance decline on the target user group $\mathcal{U}_A$, indicating that excessive KL regularization hinders the model from adapting sufficiently to new user preferences. Conversely, when $\lambda$ is too small, the performance on the overall user set $\mathcal{U}$ deteriorates, suggesting that the model suffers from forgetting previously learned preferences. These results demonstrate that an appropriate balance is necessary. Empirically, setting $\lambda = 2 \times 10^{-4}$ achieves a good trade-off between maintaining performance on $\mathcal{U}_A$ and preserving knowledge on $\mathcal{U}$.

We further analyze the impact of the threshold parameter $t$ on EvoRec's performance. As illustrated in the figure, EvoRec is sensitive to the choice of $t$. When $t$ increases from a small value, the model’s hit rate (HR) steadily improves and reaches its peak at $t = 30$, after which the performance gradually declines. We posit that when $t < 30\%$, the number of evolved parameters is insufficient to fully capture new user preferences. In this case, increasing $t$ enhances the model’s adaptability and contributes to performance gains. However, when $t > 30\%$, introducing too many parameters leads to conflicts between old and new preferences across different users, impeding further improvements. Therefore, setting $t = 30\%$ serves as a reasonable and effective configuration for achieving balanced model evolution.

\subsection{Filtering model Analysis (RQ3)}

\begin{table}[H]
\centering
\caption{Performance of EvoRec with BPR-MF as the filtering model across different $K$ values.}
\label{tab:filter_BPR}
\scalebox{0.8}{  
\resizebox{\textwidth}{!}{
\begin{tabular}{c|c|c|c|c|c|c|c|c}
\toprule
\multirow{3}{*}{\centering $K$} & \multicolumn{4}{c|}{\textbf{Amazon Beauty}} & \multicolumn{4}{c}{\textbf{Amazon Toys}} \\
\cmidrule{2-9}
 & \multicolumn{2}{c|}{$\mathcal{U}_A$} & \multicolumn{2}{c|}{$\mathcal{U}$} & \multicolumn{2}{c|}{$\mathcal{U}_A$} & \multicolumn{2}{c}{$\mathcal{U}$} \\
\cmidrule(lr){2-3} \cmidrule(lr){4-5} \cmidrule(lr){6-7} \cmidrule(lr){8-9}
 & HR@3 & NDCG@3 & HR@3 & NDCG@3 & HR@3 & NDCG@3 & HR@3 & NDCG@3 \\
\midrule
0 & 0.4823 & 0.4639 & 0.4098 & 0.3924 & 0.4137& 0.3946& 0.3749&0.3562 \\
1 & 0.4789 & 0.4582 & 0.4038 & 0.3863 & 0.4141 & 0.3914& 0.3685& 0.3501 \\
2 & 0.4821 & 0.4638 & 0.4044 & 0.3868 & 0.4177 & 0.3957 & 0.3694&0.3515 \\
3 & 0.4793& 0.4605 & 0.4042 & 0.4161 & 0.3930 & 0.3931 & 0.3691 & 0.3508\\
4 & 0.4708 & 0.4527 & 0.3985 & 0.3829 & 0.3887 & 0.3328 & 0.3647 &  0.3475\\
\bottomrule
\end{tabular}
}
}
\end{table}

In EvoRec, the filtering model can be any lightweight recommendation model that performs well on recommendation tasks, as it enhances recommendation performance by removing outdated user interactions (i.e., SASRec—a model with only one or two layers of low-dimensional attention mechanisms—serves this purpose effectively). To demonstrate the compatibility of EvoRec with different lightweight filtering models and its robustness to the hyperparameter \( K \), we replace the filtering model with GRU4Rec and BPR-MF, and conduct a sensitivity analysis on \( K \). As shown in Table~\ref{tab:filter_BPR}, Table~\ref{tab:filter_GRU4Rec}, and Table~\ref{tab:filter_SASRec}, the experimental results highlight that EvoRec’s ability to evolve user preferences largely depends on how accurately outdated interactions are detected. Since SASRec outperforms GRU4Rec and BPR-MF in terms of recommendation performance and outdated interaction identification accuracy, it serves as a more suitable filtering model. This conclusion is further supported by the fact that using GRU4Rec as the filtering model yields significantly better average performance than BPR-MF. Moreover, we observe that performance of EvoRec remains stable as the hyperparameter \( K \) varies within a reasonable range, suggesting that EvoRec is compatible with various lightweight filtering models and exhibits strong robustness to the choice of \( K \).
\begin{table}[H]
\centering
\caption{Performance of EvoRec with GRU4Rec as the filtering model across different $K$ values.}
\label{tab:filter_GRU4Rec}
\scalebox{0.8}{  
\resizebox{\textwidth}{!}{
\begin{tabular}{c|c|c|c|c|c|c|c|c}
\toprule
\multirow{3}{*}{\centering $K$} & \multicolumn{4}{c|}{\textbf{Amazon Beauty}} & \multicolumn{4}{c}{\textbf{Amazon Toys}} \\
\cmidrule{2-9}
 & \multicolumn{2}{c|}{$\mathcal{U}_A$} & \multicolumn{2}{c|}{$\mathcal{U}$} & \multicolumn{2}{c|}{$\mathcal{U}_A$} & \multicolumn{2}{c}{$\mathcal{U}$} \\
\cmidrule(lr){2-3} \cmidrule(lr){4-5} \cmidrule(lr){6-7} \cmidrule(lr){8-9}
 & HR@3 & NDCG@3 & HR@3 & NDCG@3 & HR@3 & NDCG@3 & HR@3 & NDCG@3 \\
\midrule
0 & 0.4823 & 0.4639 & 0.4098 & 0.3924 & 0.4137& 0.3946& 0.3749&0.3562 \\
1 & 0.4856 & 0.4649 & 0.4105 & 0.3930 & 0.4208 & 0.3981& 0.3752& 0.3568 \\
2 & 0.4895 & 0.4712 & 0.4111 & 0.3935 & 0.4254 & 0.4024 & 0.3761&0.3582 \\
3 & 0.4867& 0.4672 & 0.4109 & 0.4228 & 0.3997 & 0.3998 & 0.3758 & 0.3575\\
4 & 0.4782 & 0.4594 & 0.4052 & 0.3896 & 0.3954 & 0.33953 & 0.3714 &  0.3542\\
\bottomrule
\end{tabular}
}
}
\end{table}

\begin{table}[H]
\centering
\caption{Performance of EvoRec with SASRec as the filtering model across different $K$ values.}
\label{tab:filter_SASRec}
\scalebox{0.8}{  
\resizebox{\textwidth}{!}{
\begin{tabular}{c|c|c|c|c|c|c|c|c}
\toprule
\multirow{3}{*}{\centering $K$} & \multicolumn{4}{c|}{\textbf{Amazon Beauty}} & \multicolumn{4}{c}{\textbf{Amazon Toys}} \\
\cmidrule{2-9}
 & \multicolumn{2}{c|}{$\mathcal{U}_A$} & \multicolumn{2}{c|}{$\mathcal{U}$} & \multicolumn{2}{c|}{$\mathcal{U}_A$} & \multicolumn{2}{c}{$\mathcal{U}$} \\
\cmidrule{2-9}
 & HR@3 & NDCG@3 & HR@3 & NDCG@3 & HR@3 & NDCG@3 & HR@3 & NDCG@3 \\
\midrule

0 & 0.4823 & 0.4639 & 0.4098 & 0.3924 & 0.4137& 0.3946& 0.3749&0.3562 \\
1 & 0.4867& 0.4723& 0.4114& 0.3930 & 0.4205 & 0.3985 & 0.3769 & 0.3578 \\
2 & 0.4912 & 0.4762 & 0.4157 & 0.3991 & 0.4285& 0.4089& 0.3813&0.3629 \\
3 & 0.4834 & 0.4644 & 0.4102 &  0.3926& 0.4165 & 0.3974 & 0.3758 & 0.3588 \\
4 & 0.4785 & 0.4594 & 0.3975 & 0.3882 & 0.4074 & 0.3892 & 0.3695 & 0.3504 \\
\bottomrule
\end{tabular}
}
}
\end{table}
\subsection{Efficiency Analysis (RQ5)}
\label{analysis:Efficiency}
\begin{figure}[ht]
    \centering
     \includegraphics[width=1.0\textwidth]{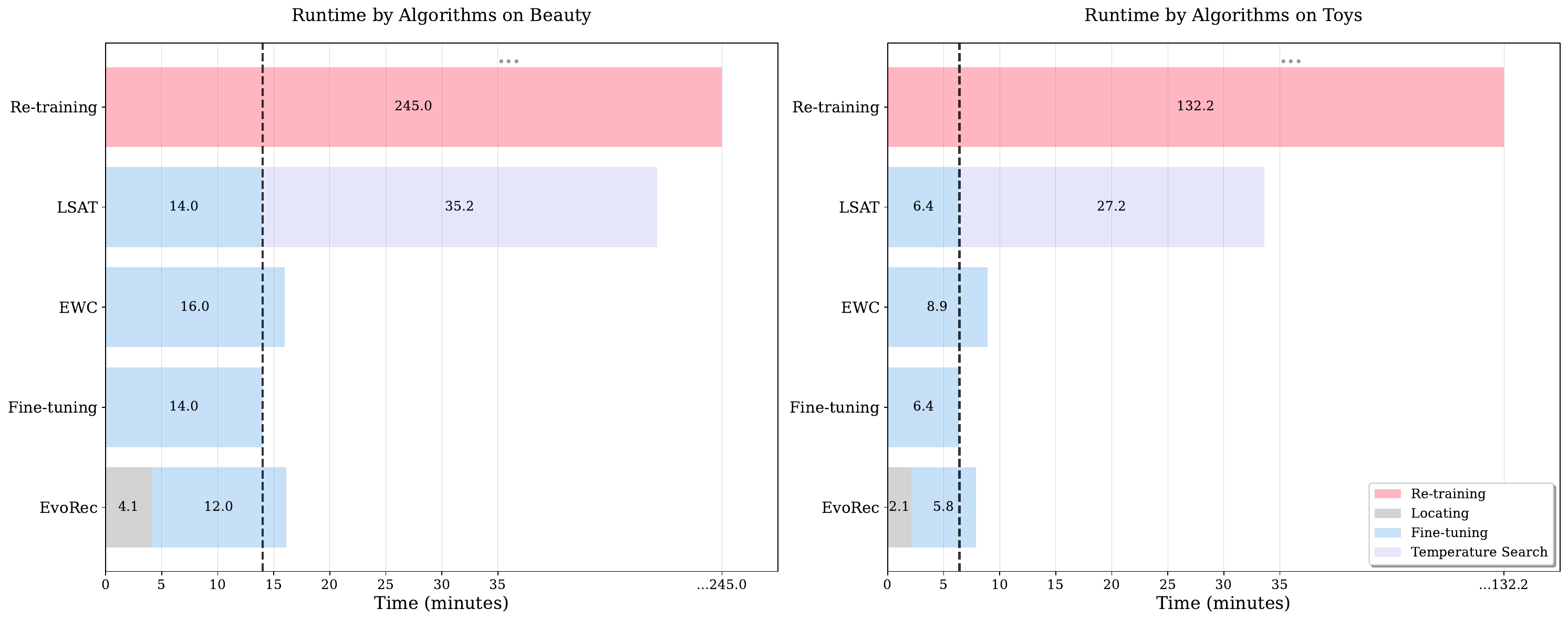}
    \caption{Time Analysis of Different Incremental Learning Paradigms on Beauty and Toys.}
    \label{fig:Time_Analysis}
\end{figure}

To assess the efficiency of different incremental learning paradigms, we analyze both computational cost and parameter update scope during the model update phase. Here, $L$ denotes the number of LLM layers, $d$ the hidden dimension, $r$ the LoRA rank, $|\mathcal{D}|$ the total number of users, $|\mathcal{D}_{\text{shift}}|$ the number of users with preference shifts, $p$ the proportion of updated LoRA parameters ($\approx 0.3$), $d_f$ the hidden dimension of the filtering model, $L_f$ its number of layers ($L_f \ll L$), $K$ the fusion search steps in LSAT, and $|\mathcal{V}|$ the validation samples in LSAT. Following the standard definition of algorithmic efficiency—which considers time and memory resources—we compare six representative paradigms: Re-training, Fine-tuning, EWC, FT-KL, LSAT, and our proposed EvoRec. The results are summarized in Table~\ref{tab:time_complexity}, Table~\ref{tab:efficiency_analysis}, and Figure~\ref{fig:Time_Analysis}.

Among all methods on Toys, Re-training incurs the highest computational cost (132.2 minutes for two epochs), as it retrains the entire model from scratch. Fine-tuning significantly reduces the update time to 6.4 minutes (3.2 minutes for one epoch) but updates the entire model without any selection mechanism, leading to potential issues such as preference forgetting. LSAT achieves improved recommendation performance by merging long- and short-term LoRA adaptations, but at the expense of a substantial increase in inference cost—resulting in a total update time of 33.6 minutes. This cost mainly comes from conducting multiple inference steps to identify the optimal merging coefficients between the long- and short-term LoRA (27.2 minutes).

In contrast, EvoRec introduces an efficient update mechanism by identifying and modifying only a small subset of sensitive parameters. As shown in Table~\ref{tab:efficiency_analysis}, EvoRec updates only 13.8M parameters—approximately 30\% of those updated in full fine-tuning—while achieving superior performance (e.g., 0.4089 NDCG@3 on Toys $\mathcal{U}_A$). The total update time for EvoRec is 7.9 minutes (2.1 minutes are spent on the Locate-Forget process), which is on par with fine-tuning and significantly faster than LSAT. Additionally, the memory overhead introduced by filtering  model of EvoRec is modest—only 886MB. We further analyze the cost associated with the filtering model in Table~\ref{tab:filtering_time}. The pRe-training of the filtering model (SASRec) takes 2.35 minutes, and its fine-tuning requires only 0.17 minutes. Crucially, since SASRec is a lightweight transformer-based model with only 1–2 layers and low-dimensional embeddings, it can be trained in parallel with the main LLM using the same data—without interfering with the LLM’s training process. This design ensures that the filtering model is always ready by the time the LLM completes initial training, thus introducing no additional delay in the incremental update pipeline.

In summary, EvoRec achieves a compelling balance between update efficiency and model effectiveness. It delivers higher performance than other incremental paradigms while maintaining comparable time and memory costs to full fine-tuning, and it significantly outperforms LSAT in computational efficiency. These results confirm EvoRec as a practical and efficient solution for incremental LLM-based recommendation.

\begin{table}[t]
\centering
\caption{Theoretical time complexity of compared methods.}
\label{tab:time_complexity}
\resizebox{\linewidth}{!}{
\begin{tabular}{lcl}
\toprule
Method & Time Complexity & \multicolumn{1}{p{5.8cm}}{\centering Key Components} \\
\midrule
Fine-tuning & $O(L \cdot |\mathcal{D}_{\text{shift}}| \cdot r \cdot d)$ & Full LoRA FT, new data \\
Re-training & $O(L \cdot |\mathcal{D}| \cdot r \cdot d)$ & Full LoRA FT, all data \\
LSAT & $O(L \cdot |\mathcal{D}_{\text{shift}}| \cdot r \cdot d + K \cdot |\mathcal{V}|)$ & Short-term LoRA + fusion search \\
EWC & $O(L \cdot |\mathcal{D}_{\text{shift}}| \cdot r \cdot d)$ & LoRA FT + EWC reg. \\
FT+KL & $O(L \cdot |\mathcal{D}_{\text{shift}}| \cdot r \cdot d)$ & LoRA FT + KL reg. \\
EvoRec (Ours) & $O\!\left(|\mathcal{D}_{\text{shift}}| \cdot (L_f \cdot d_f + L \cdot p \cdot r \cdot d)\right)$
  & Sensitive-layer LoRA ($p\!\approx\!0.3$) + light filter model ($L_f\!\ll\!L$
) \\
\bottomrule
\end{tabular}
}
\end{table}

\begin{table}[H]
\centering
\caption{Efficiency comparison of four incremental learning paradigms on the Toys dataset. Each method runs for two epochs.}
\label{tab:efficiency_analysis}
\resizebox{\linewidth}{!}{
\setlength{\tabcolsep}{4pt}
\begin{tabular}{l|c|c|c}
\toprule
\textbf{TALLRec} & \textbf{Total Update Time (min)} & \textbf{Updated Parameters} & \textbf{NDCG@3 (TOYS $\mathcal{U}_A$)} \\
\midrule
Re-training & $66.6 \times 2 = 132.2$ & 40,370,176 & 0.3948 \\
Fine-tuning & $3.2 \times 2 = 6.4$ & 40,370,176 & 0.3616 \\
LSAT & $3.2 \times 2 + 27.2 = 33.6$ & 40,370,176 & 0.3734 \\
EvoRec & $2.9 \times 2 + 2.1 = 7.9$ & 12,976,948 + 862,720 = 13,839,668 & \textbf{0.4089} \\
\bottomrule
\end{tabular}
}
\end{table}

\begin{table}[H]
\centering
\caption{Training and fine-tuning time of the filtering model used in EvoRec. The filtering model runs in parallel with the pRe-training of the main LLM.}
\label{tab:filtering_time}
\begin{tabular}{l|c}
\toprule
\textbf{Filtering Phase} & \textbf{Time (min)} \\
\midrule
Pre-train Filtering Model (in parallel with LLM) & 2.35 \\
Fine-tune Filtering Model & 0.17 \\
\midrule
\textbf{Total} & 2.52 \\
\bottomrule
\end{tabular}
\end{table}

\section{Conclusion}

This paper introduces EvoRec, an innovative framework for evolving user preferences in Large Language Model (LLM)-based recommender systems. EvoRec tackles the challenge of adapting to dynamic user interests over time without the high computational cost of traditional model retraining. By leveraging the Locate-Forget-Update paradigm, EvoRec efficiently identifies and updates only the most relevant model parameters, focusing on layers sensitive to user preference shifts. This selective updating reduces unnecessary changes to the model, preventing user preference forgetting and ensuring computational efficiency. Additionally, a filtering mechanism is employed to remove outdated interactions, further enhancing the model's ability to capture the most current user behavior and improve recommendation accuracy.

Experiments on real-world datasets demonstrate that EvoRec outperforms existing methods, offering superior recommendations for users with evolving preferences while maintaining performance for stable users. Compared to full retraining, EvoRec achieves significant computational savings, making it a practical solution for large-scale applications. By focusing on evolution rather than constant updates, EvoRec provides a balanced approach to performance and efficiency, laying the groundwork for future research in adaptive recommendation systems. Future work will aim to refine EvoRec and explore its integration with other advanced models to enhance robustness and scalability.

\begin{acks}
This research was supported by the National Science and Technology Major Project (No. 2023ZD0121\\103).
\end{acks}

\bibliographystyle{ACM-Reference-Format}
\bibliography{sample-base}

@inproceedings{markov,
  title={Factorizing personalized markov chains for next-basket recommendation},
  author={Rendle, Steffen and Freudenthaler, Christoph and Schmidt-Thieme, Lars},
  booktitle={Proceedings of the ACM International World Wide Web Conferences},
  pages={811--820},
  year={2010}
}

@inproceedings{GRU4Rec,
    title        = {Session-based Recommendations with Recurrent Neural Networks},
    author       = {Bal{\'{a}}zs Hidasi and
                  Alexandros Karatzoglou and
                  Linas Baltrunas and
                  Domonkos Tikk},
  
    booktitle    = {Proceedings of the International Conference on Learning Representations},
    pages={811--820},
    year = {2016}
}

@inproceedings{SASRec,
  author       = {Wang{-}Cheng Kang and
                  Julian J. McAuley},
  title        = {Self-Attentive Sequential Recommendation},
  booktitle    = {{IEEE} International Conference on Data Mining},
  pages        = {197--206},
  year         = {2018},
}

@inproceedings{CL4SRec,
  author       = {Xu Xie and
                  Fei Sun and
                  Zhaoyang Liu and
                  Shiwen Wu and
                  Jinyang Gao and
                  Jiandong Zhang and
                  Bolin Ding and
                  Bin Cui},
  title        = {Contrastive Learning for Sequential Recommendation},
  booktitle    = {Proceedings of the International Conference on Data Engineering},
  pages        = {1259--1273},
  year         = {2022}
}

@inproceedings{S3-Rec,
  author       = {Kun Zhou and
                  Hui Wang and
                  Wayne Xin Zhao and
                  Yutao Zhu and
                  Sirui Wang and
                  Fuzheng Zhang and
                  Zhongyuan Wang and
                  Ji{-}Rong Wen},
  title        = {S3-Rec: Self-Supervised Learning for Sequential Recommendation with Mutual Information Maximization},
  booktitle    = {Proceedings of the International Conference on Information and Knowledge Management},
  pages        = {1893--1902},
  year         = {2020}
}

@inproceedings{lsta,
  title={Preliminary study on incremental learning for large language model-based recommender systems},
  author={Shi, Tianhao and Zhang, Yang and Xu, Zhijian and Chen, Chong and Feng, Fuli and He, Xiangnan and Tian, Qi},
  booktitle={Proceedings of the 33rd ACM International Conference on Information and Knowledge Management},
  pages={4051--4055},
  year={2024}
}

@inproceedings{zhang2020retrain,
  title={How to retrain recommender system? A sequential meta-learning method},
  author={Zhang, Yang and Feng, Fuli and Wang, Chenxu and He, Xiangnan and Wang, Meng and Li, Yan and Zhang, Yongdong},
  booktitle={Proceedings of the 43rd International ACM SIGIR Conference on Research and Development in Information Retrieval},
  pages={1479--1488},
  year={2020}
}

@inproceedings{sml,
  title={How to retrain recommender system? A sequential meta-learning method},
  author={Zhang, Yang and Feng, Fuli and Wang, Chenxu and He, Xiangnan and Wang, Meng and Li, Yan and Zhang, Yongdong},
  booktitle={Proceedings of the 43rd International ACM SIGIR Conference on Research and Development in Information Retrieval},
  pages={1479--1488},
  year={2020}
}

@inproceedings{seq_1,
  title={BERT4Rec: Sequential recommendation with bidirectional encoder representations from transformer},
  author={Sun, Fei and Liu, Jun and Wu, Jian and Pei, Changhua and Lin, Xiao and Ou, Wenwu and Jiang, Peng},
  booktitle={Proceedings of the ACM international conference on information and knowledge management},
  pages={1441--1450},
  year={2019}
}

@inproceedings{seq_2,
  title={Contrastive learning for representation degeneration problem in sequential recommendation},
  author={Qiu, Ruihong and Huang, Zi and Yin, Hongzhi and Wang, Zijian},
  booktitle={Proceedings of the fifteenth ACM international conference on web search and data mining},
  pages={813--823},
  year={2022}
}

@inproceedings{seq_3,
  title={Progressive Self-Attention Network with Unsymmetrical Positional Encoding for Sequential Recommendation},
  author={Zhu, Yuehua and Huang, Bo and Jiang, Shaohua and Yang, Muli and Yang, Yanhua and Zhong, Wenliang},
  booktitle={Proceedings of the international ACM SIGIR conference on research and development in information retrieval},
  pages={2029--2033},
  year={2022}
}

@inproceedings{seq_4,
  title={Personalized top-n sequential recommendation via convolutional sequence embedding},
  author={Tang, Jiaxi and Wang, Ke},
  booktitle={Proceedings of the eleventh ACM international conference on web search and data mining},
  pages={565--573},
  year={2018}
}

@article{ilora,
  title={Customizing language models with instance-wise lora for sequential recommendation},
  author={Kong, Xiaoyu and Wu, Jiancan and Zhang, An and Sheng, Leheng and Lin, Hui and Wang, Xiang and He, Xiangnan},
  journal={arXiv preprint arXiv:2408.10159},
  year={2024}
}

@inproceedings{bao2023tallrec,
  title={Tallrec: An effective and efficient tuning framework to align large language model with recommendation},
  author={Bao, Keqin and Zhang, Jizhi and Zhang, Yang and Wang, Wenjie and Feng, Fuli and He, Xiangnan},
  booktitle={Proceedings of the 17th ACM Conference on Recommender Systems},
  pages={1007--1014},
  year={2023}
}

@inproceedings{liao2024llara,
  title={Llara: Large language-recommendation assistant},
  author={Liao, Jiayi and Li, Sihang and Yang, Zhengyi and Wu, Jiancan and Yuan, Yancheng and Wang, Xiang and He, Xiangnan},
  booktitle={Proceedings of the 47th International ACM SIGIR Conference on Research and Development in Information Retrieval},
  pages={1785--1795},
  year={2024}
}

@inproceedings{zheng2024harnessing,
  title={Harnessing large language models for text-rich sequential recommendation},
  author={Zheng, Zhi and Chao, Wenshuo and Qiu, Zhaopeng and Zhu, Hengshu and Xiong, Hui},
  booktitle={Proceedings of the ACM on Web Conference 2024},
  pages={3207--3216},
  year={2024}
}

@article{hu2021lora,
  title={Lora: Low-rank adaptation of large language models},
  author={Hu, Edward J and Shen, Yelong and Wallis, Phillip and Allen-Zhu, Zeyuan and Li, Yuanzhi and Wang, Shean and Wang, Lu and Chen, Weizhu},
  journal={arXiv preprint arXiv:2106.09685},
  year={2021}
}

@article{AdaLoRA,
  title={AdaLoRA: Adaptive budget allocation for parameter-efficient fine-tuning},
  author={Zhang, Qingru and Chen, Minshuo and Bukharin, Alexander and Karampatziakis, Nikos and He, Pengcheng and Cheng, Yu and Chen, Weizhu and Zhao, Tuo},
  journal={arXiv preprint arXiv:2303.10512},
  year={2023}
}

@article{adapater,
  title={Towards a unified view of parameter-efficient transfer learning},
  author={He, Junxian and Zhou, Chunting and Ma, Xuezhe and Berg-Kirkpatrick, Taylor and Neubig, Graham},
  journal={arXiv preprint arXiv:2110.04366},
  year={2021}
}

@inproceedings{p5,
  title={Recommendation as language processing (rlp): A unified pretrain, personalized prompt \& predict paradigm (p5)},
  author={Geng, Shijie and Liu, Shuchang and Fu, Zuohui and Ge, Yingqiang and Zhang, Yongfeng},
  booktitle={Proceedings of the 16th ACM Conference on Recommender Systems},
  pages={299--315},
  year={2022}
}

@inproceedings{MoRec,
  title={Where to go next for recommender systems? id-vs. modality-based recommender models revisited},
  author={Yuan, Zheng and Yuan, Fajie and Song, Yu and Li, Youhua and Fu, Junchen and Yang, Fei and Pan, Yunzhu and Ni, Yongxin},
  booktitle={Proceedings of the 46th International ACM SIGIR Conference on Research and Development in Information Retrieval},
  pages={2639--2649},
  year={2023}
}

@inproceedings{llm_esr,
  title={LLM-ESR: Large Language Models Enhancement for Long-tailed Sequential Recommendation},
  author={Liu, Qidong and Wu, Xian and Wang, Yejing and Zhang, Zijian and Tian, Feng and Zheng, Yefeng and Zhao, Xiangyu},
  booktitle={The Thirty-eighth Annual Conference on Neural Information Processing Systems},
  year={2024}
}

@inproceedings{wei2024llmrec,
  title={Llmrec: Large language models with graph augmentation for recommendation},
  author={Wei, Wei and Ren, Xubin and Tang, Jiabin and Wang, Qinyong and Su, Lixin and Cheng, Suqi and Wang, Junfeng and Yin, Dawei and Huang, Chao},
  booktitle={Proceedings of the 17th ACM International Conference on Web Search and Data Mining},
  pages={806--815},
  year={2024}
}

@article{llm_survey1,
  title={A survey on large language models for recommendation},
  author={Wu, Likang and Zheng, Zhi and Qiu, Zhaopeng and Wang, Hao and Gu, Hongchao and Shen, Tingjia and Qin, Chuan and Zhu, Chen and Zhu, Hengshu and Liu, Qi and others},
  journal={World Wide Web},
  volume={27},
  number={5},
  pages={60},
  year={2024},
  publisher={Springer}
}

@article{llm_survey2,
  title={Large language models for generative recommendation: A survey and visionary discussions},
  author={Li, Lei and Zhang, Yongfeng and Liu, Dugang and Chen, Li},
  journal={arXiv preprint arXiv:2309.01157},
  year={2023}
}

@inproceedings{llm_survey3,
  title={Large language models for recommendation: Progresses and future directions},
  author={Bao, Keqin and Zhang, Jizhi and Zhang, Yang and Wenjie, Wang and Feng, Fuli and He, Xiangnan},
  booktitle={Proceedings of the Annual International ACM SIGIR Conference on Research and Development in Information Retrieval in the Asia Pacific Region},
  pages={306--309},
  year={2023}
}

@inproceedings{RNN2,
  title={Neural attentive session-based recommendation},
  author={Li, Jing and Ren, Pengjie and Chen, Zhumin and Ren, Zhaochun and Lian, Tao and Ma, Jun},
  booktitle={Proceedings of the 2017 ACM on Conference on Information and Knowledge Management},
  pages={1419--1428},
  year={2017}
}

@inproceedings{graph_1,
  author       = {Ruihong Qiu and
                  Jingjing Li and
                  Zi Huang and
                  Hongzhi Yin},
  title        = {Rethinking the Item Order in Session-based Recommendation with Graph
                  Neural Networks},
  booktitle    = {Proceedings of the International Conference on Information
                  and Knowledge Management},
  pages        = {579--588},
  year         = {2019}
}

@inproceedings{graph_2,
  title={Sequential recommendation with graph neural networks},
  author={Chang, Jianxin and Gao, Chen and Zheng, Yu and Hui, Yiqun and Niu, Yanan and Song, Yang and Jin, Depeng and Li, Yong},
  booktitle={Proceedings of the 44th international ACM SIGIR conference on research and development in information retrieval},
  pages={378--387},
  year={2021}
}

@article{graph_3,
  author       = {Ruihong Qiu and
                  Zi Huang and
                  Tong Chen and
                  Hongzhi Yin},
  title        = {Exploiting Positional Information for Session-Based Recommendation},
  journal      = {{ACM} Transactions on Information Systems},
  volume       = {40},
  number       = {2},
  pages        = {35:1--35:24},
  year         = {2022}
}

@inproceedings{graph_4,
  author       = {Ruihong Qiu and
                  Hongzhi Yin and
                  Zi Huang and
                  Tong Chen},
  title        = {{GAG:} Global Attributed Graph Neural Network for Streaming Session-based
                  Recommendation},
  booktitle    = {Proceedings of the International conference on
                  research and development in Information Retrieval},
  pages        = {669--678},
  year         = {2020}
}

@inproceedings{fmlp,
  title={Filter-enhanced MLP is all you need for sequential recommendation},
  author={Zhou, Kun and Yu, Hui and Zhao, Wayne Xin and Wen, Ji-Rong},
  booktitle={Proceedings of the ACM web conference 2022},
  pages={2388--2399},
  year={2022}
}

@inproceedings{BPR_MF,
author = {Rendle, Steffen and Freudenthaler, Christoph and Gantner, Zeno and Schmidt-Thieme, Lars},
title = {BPR: Bayesian personalized ranking from implicit feedback},
year = {2009},
isbn = {9780974903958},
publisher = {AUAI Press},
address = {Arlington, Virginia, USA},
booktitle = {Proceedings of the Twenty-Fifth Conference on Uncertainty in Artificial Intelligence},
pages = {452–461},
numpages = {10},
location = {Montreal, Quebec, Canada},
series = {UAI '09}
}

@article{ewc,
  title={Overcoming catastrophic forgetting in neural networks},
  author={Kirkpatrick, James and Pascanu, Razvan and Rabinowitz, Neil and Veness, Joel and Desjardins, Guillaume and Rusu, Andrei A and Milan, Kieran and Quan, John and Ramalho, Tiago and Grabska-Barwinska, Agnieszka and others},
  journal={Proceedings of the national academy of sciences},
  volume={114},
  number={13},
  pages={3521--3526},
  year={2017},
  publisher={National Acad Sciences}
}

@inproceedings{FPMC,
  title={Factorizing personalized markov chains for next-basket recommendation},
  author={Rendle, Steffen and Freudenthaler, Christoph and Schmidt-Thieme, Lars},
  booktitle={Proceedings of the 19th international conference on World wide web},
  pages={811--820},
  year={2010}
}

@inproceedings{bert4Rec,
  title={BERT4Rec: Sequential recommendation with bidirectional encoder representations from transformer},
  author={Sun, Fei and Liu, Jun and Wu, Jian and Pei, Changhua and Lin, Xiao and Ou, Wenwu and Jiang, Peng},
  booktitle={Proceedings of the 28th ACM international conference on information and knowledge management},
  pages={1441--1450},
  year={2019}
}

@inproceedings{start1,
  title={How Important is Periodic Model update in Recommender System?},
  author={Lee, Hyunsung and Yoo, Sungwook and Lee, Dongjun and Kim, Jaekwang},
  booktitle={Proceedings of the 46th International ACM SIGIR Conference on Research and Development in Information Retrieval},
  pages={2661--2668},
  year={2023}
}

@inproceedings{fine-tuneing1,
  title={Online-updating regularized kernel matrix factorization models for large-scale recommender systems},
  author={Rendle, Steffen and Schmidt-Thieme, Lars},
  booktitle={Proceedings of the 2008 ACM conference on Recommender systems},
  pages={251--258},
  year={2008}
}

@inproceedings{fine-tuneing2,
  title={Neural memory streaming recommender networks with adversarial training},
  author={Wang, Qinyong and Yin, Hongzhi and Hu, Zhiting and Lian, Defu and Wang, Hao and Huang, Zi},
  booktitle={Proceedings of the 24th ACM SIGKDD International Conference on Knowledge Discovery \& Data Mining},
  pages={2467--2475},
  year={2018}
}

@inproceedings{sample1,
  title={Real-time top-n recommendation in social streams},
  author={Diaz-Aviles, Ernesto and Drumond, Lucas and Schmidt-Thieme, Lars and Nejdl, Wolfgang},
  booktitle={Proceedings of the sixth ACM conference on Recommender systems},
  pages={59--66},
  year={2012}
}

@inproceedings{sample2,
  title={Streaming ranking based recommender systems},
  author={Wang, Weiqing and Yin, Hongzhi and Huang, Zi and Wang, Qinyong and Du, Xingzhong and Nguyen, Quoc Viet Hung},
  booktitle={The 41st International ACM SIGIR Conference on Research \& Development in Information Retrieval},
  pages={525--534},
  year={2018}
}

@inproceedings{meta1,
  title={Long short-term temporal meta-learning in online recommendation},
  author={Xie, Ruobing and Wang, Yalong and Wang, Rui and Lu, Yuanfu and Zou, Yuanhang and Xia, Feng and Lin, Leyu},
  booktitle={Proceedings of the Fifteenth ACM International Conference on Web Search and Data Mining},
  pages={1168--1176},
  year={2022}
}

@inproceedings{LoRAMoE,
  title={LoRAMoE: Alleviating world knowledge forgetting in large language models via MoE-style plugin},
  author={Dou, Shihan and Zhou, Enyu and Liu, Yan and Gao, Songyang and Shen, Wei and Xiong, Limao and Zhou, Yuhao and Wang, Xiao and Xi, Zhiheng and Fan, Xiaoran and others},
  booktitle={Proceedings of the 62nd Annual Meeting of the Association for Computational Linguistics (Volume 1: Long Papers)},
  pages={1932--1945},
  year={2024}
}

@inproceedings{alphafuse,
author = {Hu, Guoqing and Zhang, An and Liu, Shuo and Cai, Zhibo and Yang, Xun and Wang, Xiang},
title = {AlphaFuse: Learn ID Embeddings for Sequential Recommendation in Null Space of Language Embeddings},
year = {2025},
isbn = {9798400715921},
publisher = {Association for Computing Machinery},
address = {New York, NY, USA},
url = {https://doi.org/10.1145/3726302.3729894},
doi = {10.1145/3726302.3729894},
booktitle = {Proceedings of the 48th International ACM SIGIR Conference on Research and Development in Information Retrieval},
pages = {1614–1623},
numpages = {10},
location = {Padua, Italy},
series = {SIGIR '25}
}

@article{FT-KL,
  title={Locating and editing factual associations in gpt},
  author={Meng, Kevin and Bau, David and Andonian, Alex and Belinkov, Yonatan},
  journal={Advances in neural information processing systems},
  volume={35},
  pages={17359--17372},
  year={2022}
}

@article{guo_1,
  title={Federated Semantic Learning for Privacy-preserving Cross-domain Recommendation},
  author={Lu, Ziang and Guo, Lei and Yu, Xu and Cheng, Zhiyong and Han, Xiaohui and Zhu, Lei},
  journal={ACM Transactions on Information Systems},
  year={2025},
  publisher={ACM New York, NY}
}

@article{guo_2,
  title={Semantic-enhanced Co-attention Prompt Learning for Non-overlapping Cross-Domain Recommendation},
  author={Guo, Lei and Song, Chenlong and Guo, Feng and Han, Xiaohui and Chang, Xiaojun and Zhu, Lei},
  journal={arXiv preprint arXiv:2505.19085},
  year={2025}
}

@inproceedings{guo_3,
  title={Prompt-enhanced federated content representation learning for cross-domain recommendation},
  author={Guo, Lei and Lu, Ziang and Yu, Junliang and Nguyen, Quoc Viet Hung and Yin, Hongzhi},
  booktitle={Proceedings of the ACM Web Conference 2024},
  pages={3139--3149},
  year={2024}
}

@inproceedings{liao_1,
  title={Mitigating Distribution Shifts in Sequential Recommendation: An Invariance Perspective},
  author={Liao, Yuxin and Yang, Yonghui and Hou, Min and Wu, Le and Xu, Hefei and Liu, Hao},
  booktitle={Proceedings of the 48th International ACM SIGIR Conference on Research and Development in Information Retrieval},
  pages={1603--1613},
  year={2025}
}

@inproceedings{hu_1,
  title={Alphafuse: Learn id embeddings for sequential recommendation in null space of language embeddings},
  author={Hu, Guoqing and Zhang, An and Liu, Shuo and Cai, Zhibo and Yang, Xun and Wang, Xiang},
  booktitle={Proceedings of the 48th International ACM SIGIR Conference on Research and Development in Information Retrieval},
  pages={1614--1623},
  year={2025}
}

@inproceedings{dang_1,
  title={Data augmentation as free lunch: Exploring the test-time augmentation for sequential recommendation},
  author={Dang, Yizhou and Liu, Yuting and Yang, Enneng and Huang, Minhan and Guo, Guibing and Zhao, Jianzhe and Wang, Xingwei},
  booktitle={Proceedings of the 48th International ACM SIGIR Conference on Research and Development in Information Retrieval},
  pages={1466--1475},
  year={2025}
}

@article{qwen3,
  title={Qwen3 technical report},
  author={Yang, An and Li, Anfeng and Yang, Baosong and Zhang, Beichen and Hui, Binyuan and Zheng, Bo and Yu, Bowen and Gao, Chang and Huang, Chengen and Lv, Chenxu and others},
  journal={arXiv preprint arXiv:2505.09388},
  year={2025}
}

@article{deepseek,
  title={Deepseek-v3 technical report},
  author={Liu, Aixin and Feng, Bei and Xue, Bing and Wang, Bingxuan and Wu, Bochao and Lu, Chengda and Zhao, Chenggang and Deng, Chengqi and Zhang, Chenyu and Ruan, Chong and others},
  journal={arXiv preprint arXiv:2412.19437},
  year={2024}
}

@article{kimi,
  title={Kimi k1. 5: Scaling reinforcement learning with llms},
  author={Team, Kimi and Du, Angang and Gao, Bofei and Xing, Bowei and Jiang, Changjiu and Chen, Cheng and Li, Cheng and Xiao, Chenjun and Du, Chenzhuang and Liao, Chonghua and others},
  journal={arXiv preprint arXiv:2501.12599},
  year={2025}
}

@inproceedings{huang_1,
  title={Representation learning with large language models for recommendation},
  author={Ren, Xubin and Wei, Wei and Xia, Lianghao and Su, Lixin and Cheng, Suqi and Wang, Junfeng and Yin, Dawei and Huang, Chao},
  booktitle={Proceedings of the ACM web conference 2024},
  pages={3464--3475},
  year={2024}
}

@inproceedings{huang_2,
  title={Llmrec: Large language models with graph augmentation for recommendation},
  author={Wei, Wei and Ren, Xubin and Tang, Jiabin and Wang, Qinyong and Su, Lixin and Cheng, Suqi and Wang, Junfeng and Yin, Dawei and Huang, Chao},
  booktitle={Proceedings of the 17th ACM international conference on web search and data mining},
  pages={806--815},
  year={2024}
}

@inproceedings{longlife_1,
  title={Cmt: A memory compression method for continual knowledge learning of large language models},
  author={Li, Dongfang and Sun, Zetian and Hu, Xinshuo and Hu, Baotian and Zhang, Min},
  booktitle={Proceedings of the AAAI Conference on Artificial Intelligence},
  volume={39},
  number={23},
  pages={24413--24421},
  year={2025}
}

@article{longlife_2,
  title={Wise: Rethinking the knowledge memory for lifelong model editing of large language models},
  author={Wang, Peng and Li, Zexi and Zhang, Ningyu and Xu, Ziwen and Yao, Yunzhi and Jiang, Yong and Xie, Pengjun and Huang, Fei and Chen, Huajun},
  journal={Advances in Neural Information Processing Systems},
  volume={37},
  pages={53764--53797},
  year={2024}
}

@article{longlife_3,
  title={Alphaedit: Null-space constrained knowledge editing for language models},
  author={Fang, Junfeng and Jiang, Houcheng and Wang, Kun and Ma, Yunshan and Jie, Shi and Wang, Xiang and He, Xiangnan and Chua, Tat-Seng},
  journal={arXiv preprint arXiv:2410.02355},
  year={2024}
}

@inproceedings{o-lora,
    title = "Orthogonal Subspace Learning for Language Model Continual Learning",
    author = "Wang, Xiao  and
      Chen, Tianze  and
      Ge, Qiming  and
      Xia, Han  and
      Bao, Rong  and
      Zheng, Rui  and
      Zhang, Qi  and
      Gui, Tao  and
      Huang, Xuanjing",
    booktitle = "Findings of the Association for Computational Linguistics: EMNLP 2023",
    year = "2023",
    address = "Singapore",
    publisher = "Association for Computational Linguistics",
    pages = "10658--10671"
}

@inproceedings{cmt,
  title={Cmt: A memory compression method for continual knowledge learning of large language models},
  author={Li, Dongfang and Sun, Zetian and Hu, Xinshuo and Hu, Baotian and Zhang, Min},
  booktitle={Proceedings of the AAAI Conference on Artificial Intelligence},
  volume={39},
  number={23},
  pages={24413--24421},
  year={2025}
}

@inproceedings{kim2024,
  title={Large language models meet collaborative filtering: An efficient all-round llm-based recommender system},
  author={Kim, Sein and Kang, Hongseok and Choi, Seungyoon and Kim, Donghyun and Yang, Minchul and Park, Chanyoung},
  booktitle={Proceedings of the 30th ACM SIGKDD Conference on Knowledge Discovery and Data Mining},
  pages={1395--1406},
  year={2024}
}

@article{tois_1,
author = {Chen, Lei and Gao, Chen and Du, Xiaoyi and Luo, Hengliang and Jin, Depeng and Li, Yong and Wang, Meng},
title = {Enhancing ID-based Recommendation with Large Language Models},
year = {2025},
issue_date = {September 2025},
publisher = {Association for Computing Machinery},
address = {New York, NY, USA},
volume = {43},
number = {5},
journal = {ACM Transactions on Information Systems.},
numpages = {30}
}

@article{tois_2,
author = {Luo, Sichun and He, Bowei and Zhao, Haohan and Shao, Wei and Qi, Yanlin and Huang, Yinya and Zhou, Aojun and Yao, Yuxuan and Li, Zongpeng and Xiao, Yuanzhang and Zhan, Mingjie and Song, Linqi},
title = {RecRanker: Instruction Tuning Large Language Model as Ranker for Top-k Recommendation},
year = {2025},
issue_date = {September 2025},
publisher = {Association for Computing Machinery},
address = {New York, NY, USA},
volume = {43},
number = {5},
journal = {ACM Transactions on Information Systems.},
numpages = {31}
}










\end{document}